\documentclass[fleqn,usenatbib]{mnras}
\usepackage{amsmath}
\usepackage[T1]{fontenc}

\DeclareRobustCommand{\VAN}[3]{#2}
\let\VANthebibliography\thebibliography
\def\thebibliography{\DeclareRobustCommand{\VAN}[3]{##3}\VANthebibliography}

\usepackage{graphicx}
\usepackage{txfonts}
\usepackage{natbib}
\usepackage{url}
\usepackage{multirow}
\usepackage{amssymb}
\usepackage{mathabx}
\usepackage{siunitx}
\usepackage{slashbox}

\newcommand{\kms}{km~s$^{-1}$} 
\newcommand{\msun}{M$_{\Sun}$}
\newcommand{\e}{$\times$10}

\newcommand{\h}{H$_2$}

\newcommand{\HII}{H{\sc ii}}
\newcommand{\cd}{1\e$^{22}$~cm$^{-2}$}
\newcommand{\vect}[1]{\boldsymbol{#1}}
\newcommand{\fIRB}{$f_\text{IRB}$}

\title[Evolutionary growth of molecular clouds]{Evolutionary growth of molecular clouds as traced by their infrared bright fraction}%
\author[E. J. Watkins et al.]{
E. J. Watkins,$^{1,2}$\thanks{E-mail: elizabeth.watkins@manchester.ac.uk}
N. Peretto,$^{3}$
A. J. Rigby,$^{4}$
R. J. Smith,$^{5}$
K. Kreckel,$^{2}$
G. A. Fuller$^{1,6}$
\\
$^{1}$Jodrell Bank Centre for Astrophysics, Department of Physics and Astronomy, University of Manchester, Oxford Road, Manchester M13 9PL, UK \\
$^{2}$Astronomisches Rechen-Institut, Zentrum f\"{u}r Astronomie der Universit\"{a}t at Heidelberg, M\"{o}nchhofstraße 12-14, D-69120 Heidelberg, Germany \\
$^{3}$Cardiff Hub for Astrophysics Research \& Technology, School of Physics \& Astronomy, Cardiff University, Queens Buildings, The Parade, Cardiff CF24 3AA, UK \\
$^{4}$School of Physics and Astronomy, University of Leeds, Leeds LS2 9JT, UK \\
$^{5}$School of Physics and Astronomy, University of St Andrews, North Haugh, St Andrews, KY16 9SS \\
$^{6}$I. Physikalisches Institut, University of Cologne, Z\"ulpicher Str. 77, 50937 K\"oln, Germany\\
}

\date{Accepted XXX. Received YYY; in original form ZZZ}

\pubyear{2024}

\begin{document}
\label{firstpage}
\pagerange{\pageref{firstpage}--\pageref{lastpage}}
\maketitle

\begin{abstract} 
Understanding how stars form, evolve and impact molecular clouds is key to understanding why star formation is such an inefficient process globally. In this paper, we use the infrared bright fraction, \fIRB\ (the fraction of a given molecular cloud that appears bright against the 8~\micron\ Milky Way background) as a proxy for time evolution to test how cloud properties change as star formation evolves. We apply this metric to 12,000 high-mass star-forming molecular clouds we identify using the \textit{Herschel}--Hi-GAL survey between $|\ell|<\ang{70;;}$ on the Milky Way plane. We find clouds are not static while forming stars. Instead, molecular clouds continuously gain mass while star formation progresses. By performing principal component analysis on the cloud properties, we find that they evolve down two paths distinguished by their mass gain. Most clouds (80\%) gain four times more mass as a function of \fIRB. The remaining 20\% experience an extreme period of growth, growing in mass by a factor of 150 on average and during this period, they initially gain mass fast enough to outpace their star formation. For all clouds, it is only after half their area becomes star forming that mass loss occurs. We expect stellar feedback and potentially galactic shear is responsible. By analysing cloud positions, we suggest that the rate of mass growth may be linked to the larger galactic environment. Altogether, these results have strong implications on how we assess star forming ability on cloud scales when assuming molecular cloud masses are fixed in time.
\end{abstract}



%

\section{Introduction} 

\noindent Stars form in the small fraction of the interstellar stellar medium (ISM) that contains dense ($>$1000~cm$^{-3}$) inhomogeneous molecular structures. On galactic scales, large-scale dynamic motions and stellar feedback organise these structures into spiral arms, large galactic filaments, and voids \citep{elmegreen_star_2011,ragan_giant_2014,zucker_skeleton_2015,grisdale_impact_2017,watkins_phangsjwst_2023}. On smaller scales ($\sim$10~pc), the added influence of self-gravity and turbulence \citep{chen_self-gravitating_2020,abe_classification_2021} organises the gas into filamentary molecular clouds \citep{schneider_catalog_1979,molinari_clouds_2010} and hubs \citep{myers_filamentary_2009,anderson_alma_2021} that can contain star-forming clumps ($\sim$0.1--1~pc) and cores ($<0.1$~pc).

While the exact definition and nature of a molecular cloud varies depending on the spatial scale resolved and the tracer used, they typically represent where we expect molecular gas can evolve into discrete star forming complexes \citep{chevance_molecular_2020}.  
Stellar feedback from these complexes, and shear from larger scales disrupts molecular gas across these scales. In particular, it has been shown that stellar feedback is necessary to explain low star formation efficiencies (SFEs) in the nearby universe \citep{krumholz_radiation-hydrodynamic_2007,hopkins_galaxies_2014}. Therefore, a complete theory of star formation in molecular gas structures needs a unified picture of star formation on large and small scales \citep{dobbs_frequency_2015,pettitt_how_2020,smith_cloud_2020}. However, because molecular clouds evolve over millions of years, we cannot follow its exact evolution observationally for individual structures. We must instead piece it together statistically using a large sample of molecular clouds in different environments at different evolutionary stages. These samples allow us to estimate the timeline for a given molecular cloud by using the average properties of less and more evolved clouds as surrogates, providing we have a tracer that is able to detect the presence or absence of star formation. 
Such techniques have been employed in various forms in the Milky Way and nearby galaxies \citep{battersby_lifetimes_2017,chevance_lifecycle_2020,mazumdar_high_2021,rigby_gaston_2021,rigby_dynamic_2024}. For example \cite{battersby_lifetimes_2017} used 8~\micron\ emission to trace both the amount of gas that is currently starless or is star forming on a pixel-by-pixel basis. This allowed them to measure the fraction of high-mass pixels (as defined using Hi-GAL data) that were infrared bright or dark over $\ang{2;;}\times\ang{2;;}$ of the Milky Way plane. They found that $\sim$65\% of pixels were quiescent while $\sim35$\% were star-forming, and when linked to co-spatial objects with known lifetimes, they estimated the starless and star forming phases of high-mass pixels were 0.2--1.7~Myr and 0.1--0.7~Myr respectively.

Observations at 8~\micron\ are somewhat unique in how it traces the presence or absence of star formation in the Milky Way. The diffuse background $\sim$8~\micron\ emission across the inner $\ell<|\ang{70;;}|$ part of the Milky Way plane (with typical values of the order $\sim$100~MJy~sr$^{-1}$) is bright enough that it can be absorbed by foreground cold dust, appearing as dark patches against the bright emission \citep{peretto_initial_2009}. Cold dust traces quiescent starless molecular gas, therefore it can be used to trace absent star formation. At the same time, direct emission from protostellar objects is sufficiently bright to outshine the diffuse background in this band. Moreover, the stellar light (especially UV light) emitted by recently formed high-mass stars cause polycyclic aromatic hydrocarbons (PAHs) to fluoresce throughout the mid-infrared (mid-IR). This is particularly bright at $\sim$8~\micron\ since it covers multiple PAHs bands and the stellar and cold dust continuum over this range are simultaneously low \citep{marble_aromatic_2010}. Statistically, feedback from high-mass stars, which UV photons trace, marks the end of the star formation in molecular gas \citep{chevance_lifecycle_2020}, and so it can also be used to trace the progression of star formation. Therefore, we can use broadband 8~\micron\ observations to calculate a relative evolutionary timeline and status of the star formation within molecular clouds on a cloud-by-cloud basis across the Milky Way plane.

For this purpose we structure the paper as follows: Section \ref{sec:obs} introduces the data sets needed to identify molecular clouds and to determine their evolutionary stage and properties. In Section \ref{sec:catalogue}, we produce a catalogue of molecular clouds across the Milky Way plane. Section \ref{sec:firb} outlines how to estimate how evolved the star formation is in molecular clouds using mid-IR emission and tests how effective it is as an evolutionary tracer. In Section \ref{sec:cloud_props}, we present our results and trends of cloud properties as a function of evolutionary stage. In Section \ref{sec:discuss} we discuss the implications of these results in the context of cloud evolution, and present our closing remarks in Section \ref{sec:conclude}.

\section{Observations and survey catalogues} \label{sec:obs}

\subsection{Hi-GAL survey and source catalogues} \label{higal-obs}
\noindent The {\it Herschel} infrared Galactic Plane Survey (Hi-GAL) was a key project of {\it Herschel} \citep{molinari_hi-gal:_2010} to measure cold dust as a tracer of molecular gas and star formation. It mapped $\sim$1$^{\circ}$ above and below the Galactic plane over a 360$^{\circ}$ view at 70~\micron, 160~\micron, 250~\micron, 350~\micron, and 500~\micron\ with resolutions of $\ang{;;7}$, $\ang{;;12}$, $\ang{;;18}$, $\ang{;;24}$, and $\ang{;;36}$, respectively. In addition to providing individual $\sim$2.2 deg$^2$ maps of the Milky Way plane, the Hi-GAL team has produced a compact source catalogue for each waveband, each containing $\sim$100,000 entries \citep{molinari_hi-gal_2016}. Each catalogue currently covers $|\ell| \lesssim 70^{\circ}$ and $|b| \le 1^{\circ}$. Sources were extracted using the CuTEx algorithm \citep{molinari_source_2011}, which is able to detect sources over a large range of background conditions (both low and high background values). 

\subsection{GLIMPSE I/II}
\noindent The Galactic Legacy Infrared Mid-Plane Survey Extraordinaire (GLIMPSE) \citep{churchwell_spitzer/glimpse_2009} was a near-to-mid infrared Galactic plane survey taken with the InfraRed Array Camera (IRAC) on-board the {\it Spitzer} Space Telescope \citep{fazio_infrared_2004,werner_spitzer_2004}. The survey covered Galactic longitudes of $|\ell|\leq65^{\circ}$ and Galactic latitudes $|b|\leq1^{\circ}$ (extending to 1.5$^{\circ}$ between $|\ell|$=5--2$^{\circ}$ and extending to 2$^{\circ}$ for longitudes $|\ell|\leq2^{\circ}$). The survey imaged the sky in 3.6, 4.5, 5.8, and 8.0~$\mu$m. 8.0~$\mu$m traces both internal heating from protostellar objects and traces \HII\ regions via the fluorescence of PAHs, caused by UV emission. The angular resolution, pixel resolution and the average rms uncertainty are $\ang{;;1.9}$, $\ang{;;1.2}$ and 0.3~MJy~sr$^{-1}$ respectively at 8.0~$\mu$m for this survey. For this study, we use the $8.0\,\mu{\rm m}$ flux density to indicate how much of a given cloud is currently forming stars.
 
\section{Molecular cloud catalogue and properties} \label{sec:catalogue}

\subsection{Cloud identification} \label{sec:cloud_id}
To trace molecular gas, we use dust emission from the Hi-GAL survey between $|\ell|\leq70^{\circ}$ $|b|\leq1^\circ$. We chose these extents to match the survey extent of GLIMPSE, which we use to trace how evolved the star formation is in Section \ref{sec:firb}. For Hi-GAL data, we are limited to a 2D projected view, which makes foreground and background emission indistinguishable. As a result, the amount of molecular gas we measure will be overestimated, and the structures we detect will be more extended due to the increased column of gas along the line of sight. To limit these impacts, we use the $\ang{;;18}$ resolution background subtracted \h\ column density maps derived from Hi-GAL observations from \cite{peretto_initial_2016}. The background has been defined using a median filter that is $\ang{;10;}$ in size (matching the largest \textit{Spitzer} dark clouds found in the GLIMPSE survey), which removes extended emission on scales larger than $\ang{;10;}$. This also sets the maximum size scale of molecular structures that we identify. 
In general, molecular clouds are nebulously defined objects due to the multi-scalar and continuous nature of the ISM. 
As a result, their appearance is not only dependent on the tracer used, but also the spatial scales being probed. Multiple detection approaches (which lead to different structures being identified as molecular clouds) are therefore needed and have been developed to detect molecular clouds \citep{williams_determining_1994,rosolowsky_bias-free_2006,rosolowsky_structural_2008,colombo_graph-based_2015,berry_fellwalkerclump_2015,rani_identification_2023}.

\begin{figure*}
	\centering
		\includegraphics[width = \textwidth]{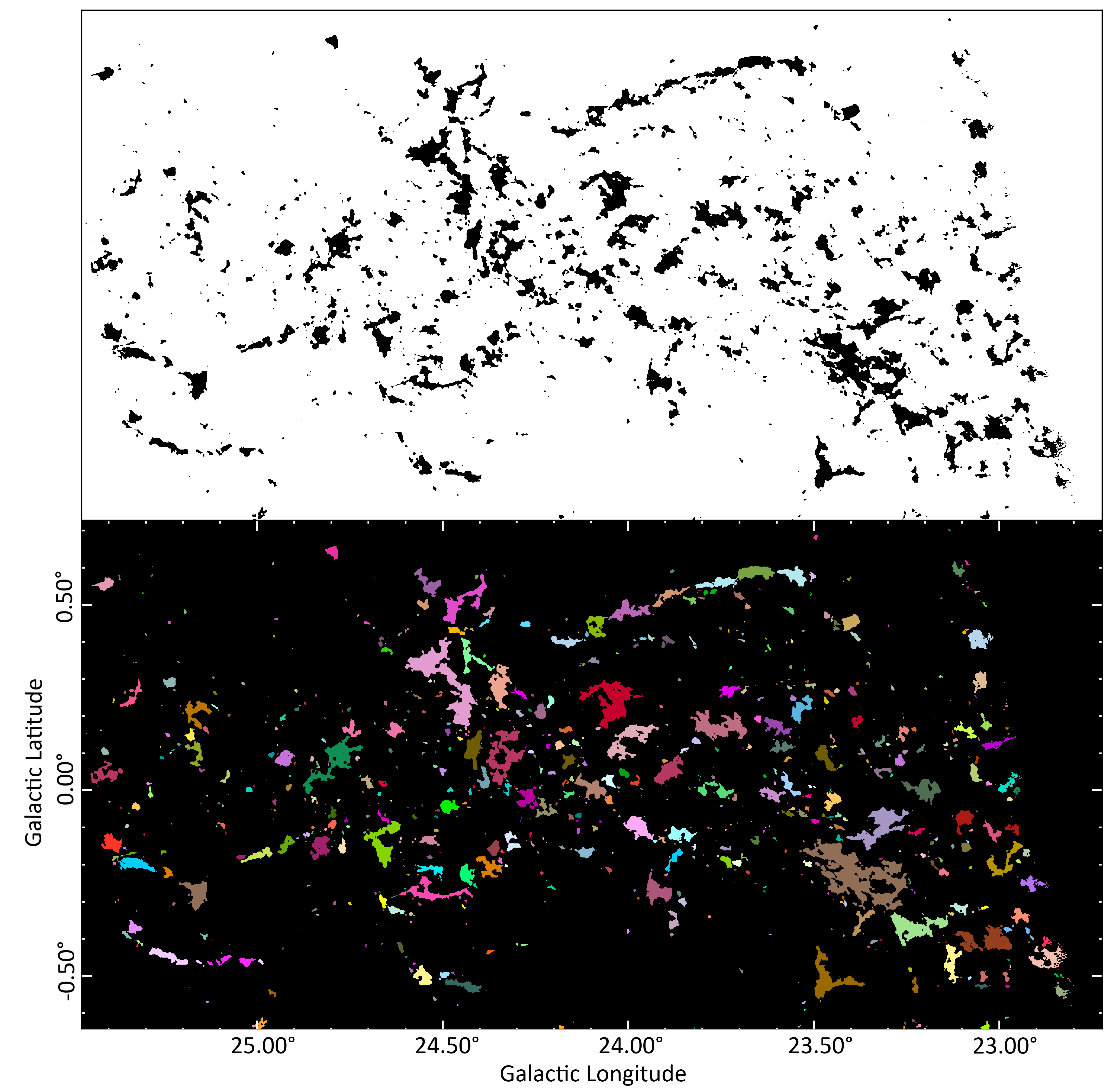}
		\caption{Molecular structures identified in \textit{Herschel} column density data between a galactic longitude of 23--25.5$^\circ$. \textit{Top} panel shows all pixels $\geq$1\e$^{22}$~cm$^{-2}$ (solid black colour), where each pixel is $\ang{;;4.5}$ in size. \textit{Bottom} panel shows the regions that are identified. Colours indicates the ID's of clouds. 
        }
	\label{fig:24_upper}
\end{figure*} 

To minimise the impact of different density thresholds have on structures, we identify clouds using any closed contour above a single column density threshold. For this threshold, we use 1\e$^{22}$~cm$^{-2}$. While 1\e$^{22}$~cm$^{-2}$ might be considered arbitrary and high -- especially considering that gas is expected to be molecular and star forming an order of magnitude below this value \citep{clark_column_2014} -- it allows us to focus on objects that are close to or beyond the predicted column density threshold needed star formation \citep{johnstone_extinction_2004,andre_filamentary_2014,priestley_neath_2023}. Finally, a high column density threshold assists with the evolution tracer, discussed further in Sect.~\ref{sec:contrast}.

\begin{figure*}
	\centering
		\includegraphics[width = \textwidth]{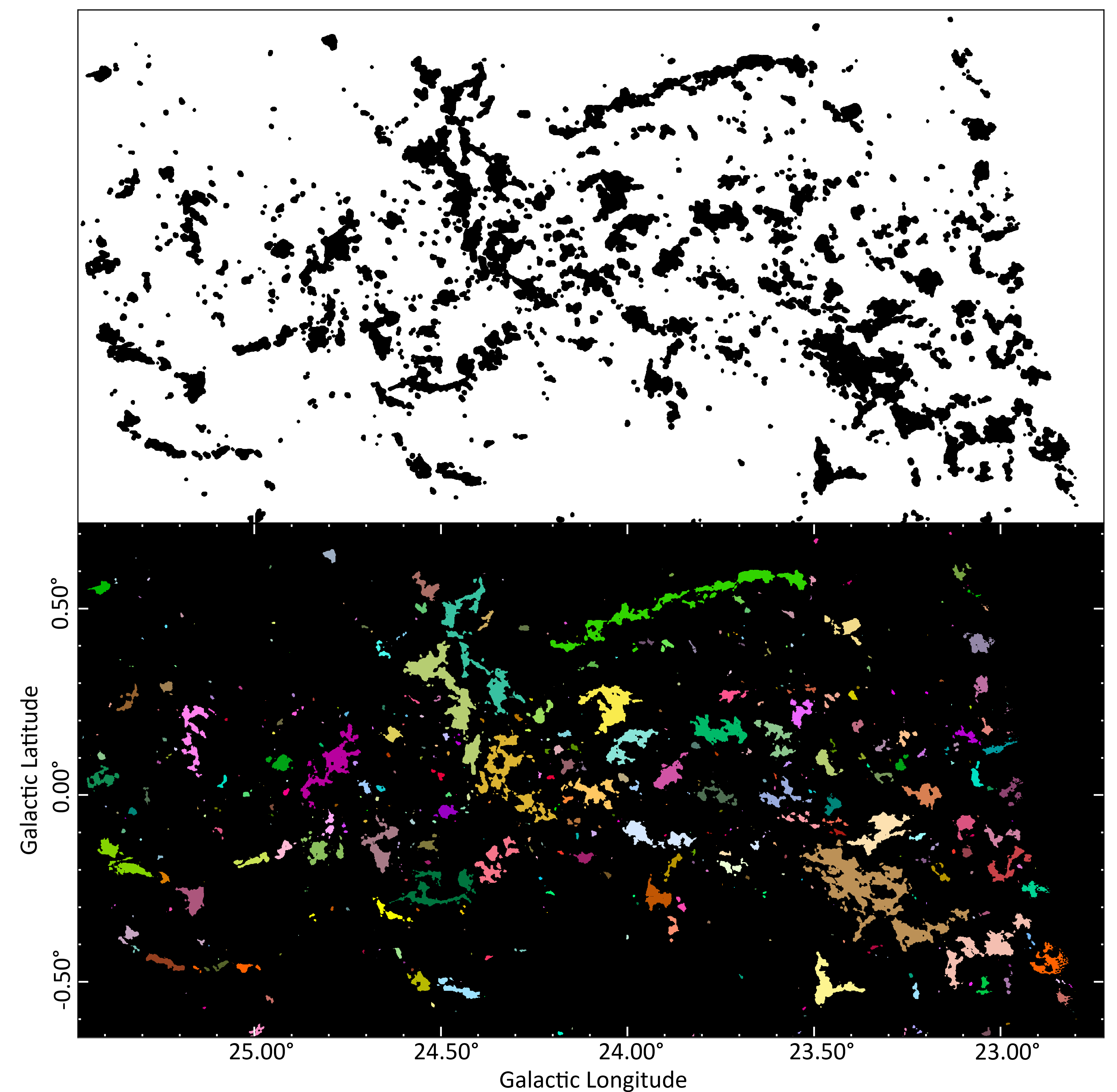}
		\caption{Applying watershed-like algorithm to connect spatial coherent substructures in Fig.~\ref{fig:24_upper}. \textit{Top} panel illustrates how substructures are connected by convolving the masked data to 21$''$. \textit{Bottom} panel shows the large scale structures identified after the watershed-like algorithm. The colours indicates the ID's of each cloud. See Fig.~\ref{fig:24_upper} to compare how substructures have been grouped together.}
	\label{fig:24_lower}
\end{figure*}

\begin{figure}
	\centering
		\includegraphics[width = \columnwidth]{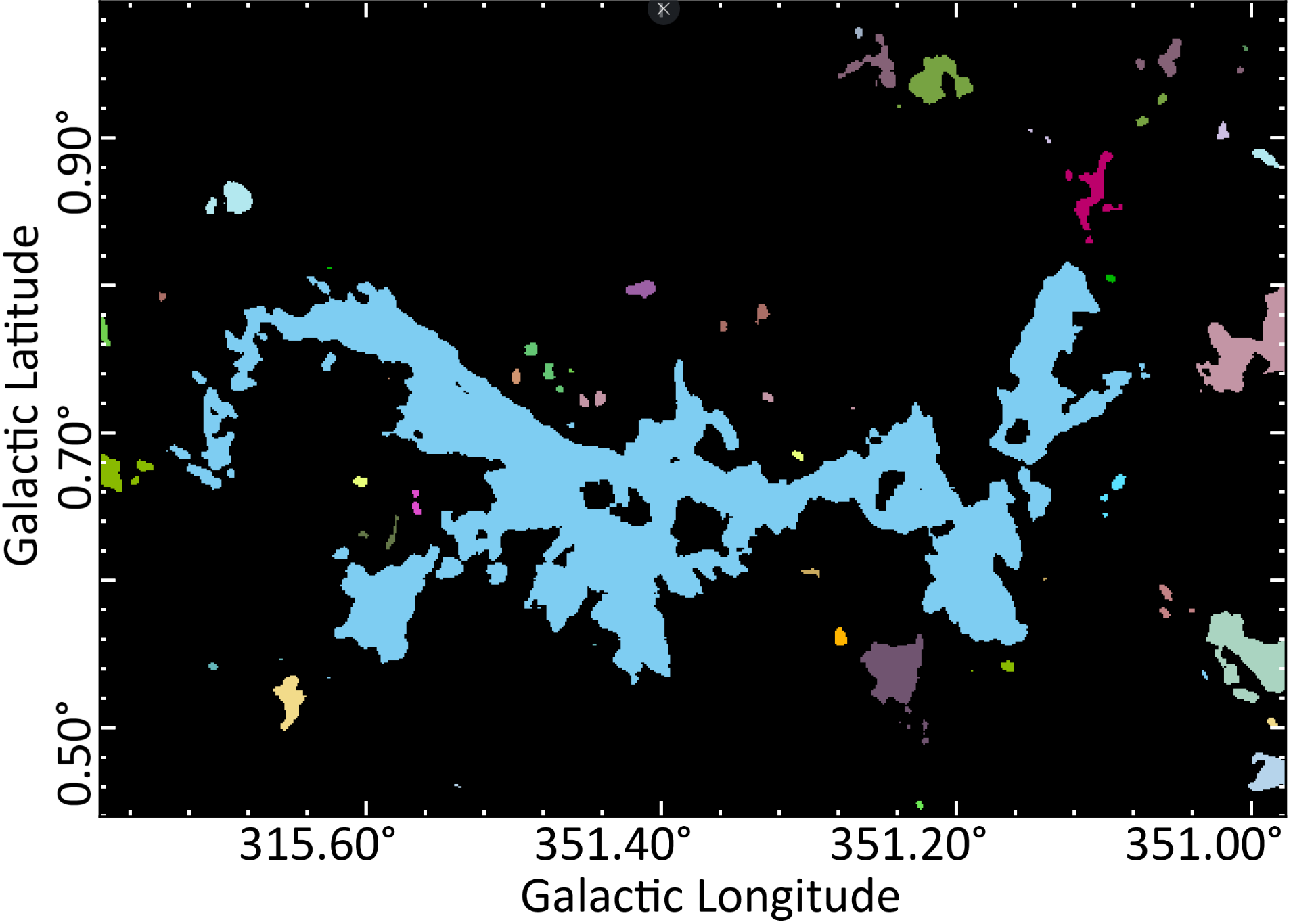}
		\caption{NGC6334 after connecting masks using watershed-like method. Light blue mask in the centre of the image shows NGC6334 where the majority of unconnected masks are merged (and so appear light blue) after performing a watershed-like method. Masks that are not light blue represent independent molecular structures identified in Sect.~\ref{sec:cloud_id}.} 
	\label{fig:NGC6334}
\end{figure}

The main disadvantage of a high column density threshold is the artificial segmentation of spatially (and velocity) coherent molecular clouds into smaller substructures (analogous to molecular `clumps'). Indeed, any column density cutoff results in arbitrary cloud boundaries within a spatially coherent structure at some larger scale due the multi-scalar and continuous nature of the ISM. However, by using median-filter-subtracted column density maps, we have already introduced a segmentation scale that limits the maximum size structures can be. Since this size scale is based on GLIMPSE observations, we prioritise studying objects closer to this spatial scale. To do this without lowering the column density threshold, we perform a watershed-like algorithm to connect spatially coherent substructures. Substructures are connected together by generating a binary map equal to 1 for pixels above a column density threshold of 1\e$^{22}$~cm$^{-2}$ and dilating these masks outward by convolving the map with a small Gaussian kernel (see example top panel of Fig.~\ref{fig:24_upper}). Any substructures that now touch are considered to be one structure, as indicated on Fig.~\ref{fig:24_lower}. To avoid structures that span size scales significantly $>\ang{;10;}$, we only convolve the masks to $\ang{;;21}$. Finally, any structures with an effective area smaller than the resolution of the column density map ($\ang{;;18}$) are removed. After visual inspection, we find it was able to join most of the gas in known star forming regions (e.g., NGC6334, see Fig.~\ref{fig:NGC6334} and Fig.~\ref{fig:l30}), without resulting in degree-sized over-merged structures everywhere, even towards crowded locations closer to the galactic centre. However due to the high column densities found specifically between $|\ell|<\ang{2.5;;}$, our algorithm always created over-merged objects (see Fig.~\ref{fig:l2}) with extremely large spatial extents ($>\ang{1;;}$). As a result, we exclude all structures between $|\ell|<\ang{2.5;;}$ from further analysis.

Before merging we identify 22851 structures and after we identify 16343. Of these, 553 are outside the exact GLIMPSE survey area so after removing these, we are left with 15790. For the rest of this paper, to simplify discussions, we refer to these grouped structures as molecular clouds (or clouds for short).

\subsection{Assigning distances to molecular clouds} \label{sec:dist}

\noindent The Hi-GAL team calculated the distance catalogue to $>$150,000 Hi-GAL sources \citep{mege_distance_2021}. These distances were assigned by cross examining all available distance tracers for each object. If maser parallax measurements were available, the source distance could be determined with a high degree of accuracy. However, only 93 sources in their catalogue had maser parallax measurements. Therefore the majority of distances in \cite{mege_distance_2021} were assigned using radial velocity measurements from molecular line or from H{\sc i} surveys with a Galactic rotation model. To resolve the near-far kinematic distance ambiguity for each source, the distance was compared to dust extinction, H{\sc i} absorption features and its projected scale height when these options are available. For example, if a source was seen in absorption, or its distance from the mid-plane at the far distance was significantly greater than the scale height of the Milky Way, then it was unlikely to be at the far distance and so it was assigned the near distance value.

Using the Hi-GAL distance catalogue, we assign a distance to each cloud only if their area masked in the previous section directly overlaps with the compact Hi-GAL source. However, complications arise for clouds (and even the initial pre-merged sources) that contain multiple Hi-GAL sources. We find the distance (and velocity) values between sources can disagree due to measurement uncertainties, source crowding along the line of sight or in some instances, clouds contain sources with identical velocities, but their distances are near-far confused. To assign a single distance to a cloud while accounting for these complex situations, we create a decision tree. In summary, the decision tree determines distances and velocities for clouds while eliminating small outliers, and rejecting clouds from further analysis when most of the distances and velocities are dissimilar. Removing such clouds is necessary to account for the fact we only used 2D spatial information when identifying clouds.

The first part of the decision tree passes clouds that contain only a single source and assigns that distance to the cloud. The next part of the decision tree deals with clouds that contain three or more distance values and aims to remove clouds that are unlikely to be spatially coherent in 3D space. For these clouds, we only assign the median distance (and velocity) if $>$50\% of the distances and the velocities remain after sigma clipping (we use the median absolute deviation with a 3 sigma confidence level), otherwise, we re-perform the sigma clip using the near distances only. If$>$50\% of the distances now remain, we assign the median near-distance value, while the rest do not get a distance assigned and are removed from our sample. We prioritise the near distance since most clouds (70-80\%) are located at the near distance \citep{urquhart_atlasgal_2018}. We investigate the impact this decision might have for our results in Appendix \ref{sec:checks}.

The next part of the decision tree deals with clouds where sigma clipping cannot be used. For very coherent distance and velocity values, sigma clipping causes erroneous source rejection and therefore cloud rejection when in fact, the distances of the sources are nearly identical. Therefore, when all the velocities and distances are within $\leq$2.5~\kms and $\leq$0.1~kpc of each other, the median distance is assigned without performing sigma clipping. We chose these limits by inspecting the velocity and distance values present when sigma clipping failed. Increasing these ranges did not significantly increase the number of clouds accepted; the values are similar enough that sigma clipping failed. However when larger limits are used, the distance and velocity assigned are more likely to be inconsistent with commonly used values found in the literature for known objects.

Finally, if a cloud contains only two distance values, sigma clipping cannot be used. For these clouds, the mean distance is assigned if both the velocities and distances are within $\leq$5~\kms and $\leq$0.2~kpc of each other. Otherwise, no distance is assigned and the cloud is removed. These limits are chosen to match the strict limits we set for coherent clouds (i.e., both clouds can different by $\leq$2.5~\kms and $\leq$0.1~kpc each, which doubles these values). Altogether, we are able to assign distances to 12686 clouds, (75\% of the catalogue). We note that 20 of these clouds had distances >30~kpc. These clouds are also rejected, leaving 12666 clouds, which is comparable to the number of objects identified in \citep{peretto_initial_2016}. We are therefore confident we are probing similar spatial scales as this paper.

\begin{figure*}
	\centering
		\includegraphics[width = \textwidth]{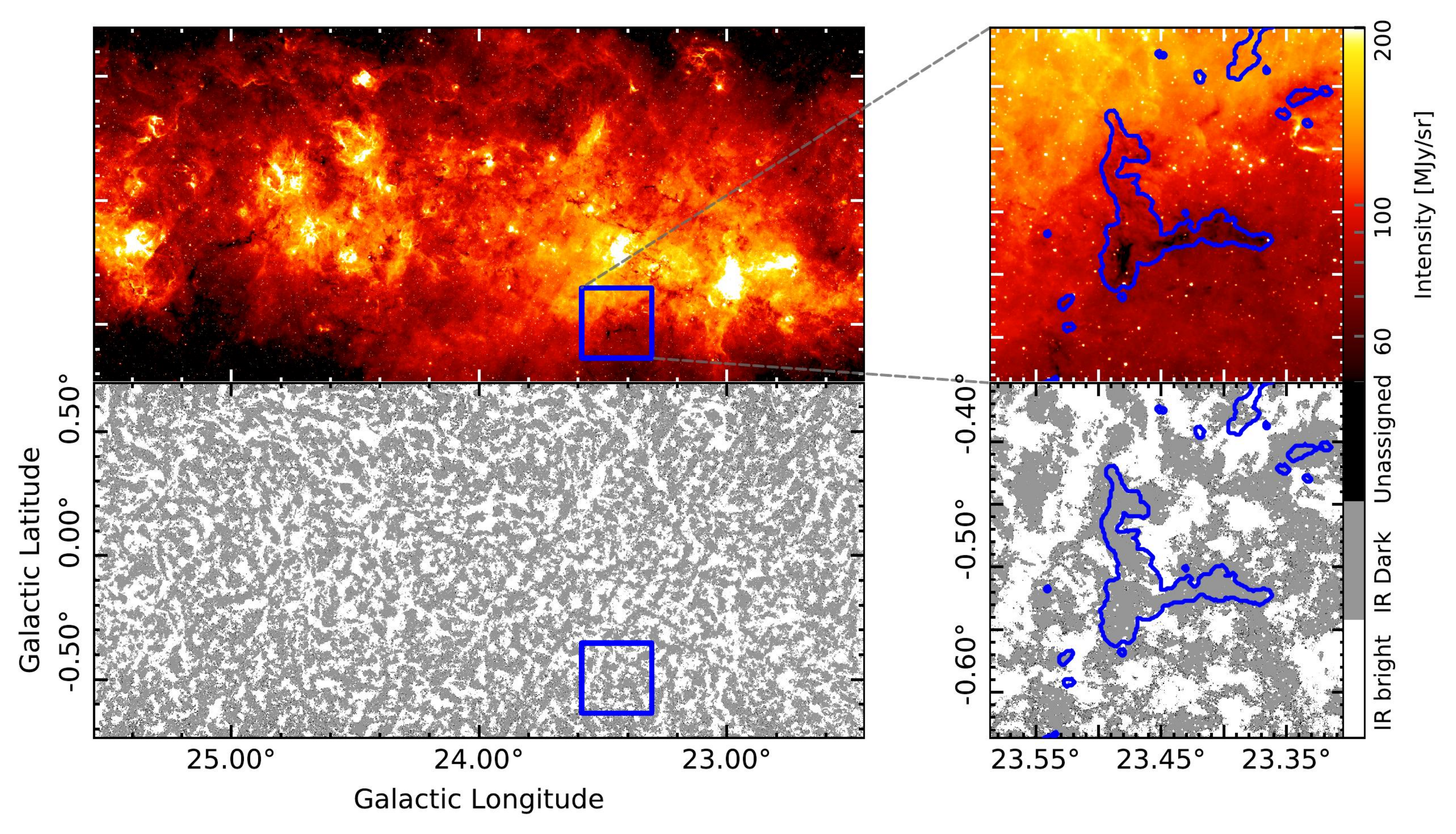}
		\caption{\emph{Spitzer} 8~\micron data and its infrared bright map derived by subtracting its background emission, along with its application to molecular clouds. \textit{Top left}: \emph{Spitzer} 8~\micron\ map from the galactic plane ($\ang{23;;}$--$\ang{25.5;;}$) showing both bright star forming, and dark quiescent molecular gas. Blue box marks the location showing an example molecular cloud with complex background emission. \textit{Top right}: Example molecular cloud seen in absorption at 8\micron\ with complex background emission. Blue contour shows the \textit{Hershel} cloud masks above a column density threshold of \cd. \textit{Bottom left}: Infrared bright map after applying a median filter of $\ang{;4.8;}$ on the GLIMPSE 8~\micron\ map and subtracting the result away from the original image. Positive pixels are deemed infrared bright (white), negative pixels are deemed infrared dark (grey) and any pixels that fall under the RMS noise (0.3~MJy~sr$^{-1}$) are left unassigned (black). \textit{bottom right}: Same example region as \textit{top right} but with the infrared bright map shown instead.}
	\label{fig:ir_map}
\end{figure*}

\section{Infrared bright fraction as an evolutionary tracer of molecular clouds} \label{sec:firb}

\noindent On the Milky Way plane, 8~\micron\ emission is able to trace both the presence and absence of star formation within individual clouds when contrasted against the bright background emission. As a result, the fraction of bright emission within the area of a cloud indicates how evolved the star formation is within that clouds. Using this fact, we define a quantity we call the infrared bright fraction (\fIRB) to represent this evolution as a fraction between zero and one. For each cloud, \fIRB\ denotes the number of bright pixels over the total number of pixels within a cloud. An \fIRB\ of 0 therefore represents the earliest stage of star formation (i.e., starless clouds) and 1 indicates star formation is underway in the entire cloud. With this definition, \fIRB\ increase as time increases. The \fIRB\ includes emission from protostellar heating, and PAHs heated by high-mass stars, both of which indicate the star formation is more evolved in the molecular gas.

To calculate \fIRB, we subtract the background emission from the GLIMPSE 8~\micron\ maps, leaving a map containing positive (bright) and negative (dark) pixel values. To model the Galactic plane background while accounting for its complex, highly varying emission (i.e., decreasing emission as a function of Galactic latitude and longitude, and local variations due to galactic structures) we use a median filter of size $\ang{;4.8;}$ for the entire GLIMPSE survey. After subtraction, we label all positive pixels above the survey RMS (0.3~MJy~sr$^{-1}$) as infrared bright with a value of one and all negative pixels below the survey RMS as infrared dark with a value of zero. Any pixels falling within the RMS noise limit remain unassigned since the pixel value could be caused by noise fluctuations. Very few pixels are left unassigned. We chose $\ang{;4.8;}$ to match the average size of the structures we identify in Sect \ref{sec:cloud_id} and because it is similar to the average sizes identified in \citep{peretto_initial_2009}.

These background subtracted infrared bright-dark maps have already been successfully applied to clumps found using New IRAM KIDs Array 2 (NIKA2) camera in \cite{rigby_gaston_2021} and clumps examined with NOEMA and ALMA in \cite{rigby_dynamic_2024} to calculate their \fIRB. In Fig.~\ref{fig:ir_map}, we illustrate the background subtracted infrared bright-dark map between $l=\ang{23;;}$--$\ang{25.5;;}$, matching the area mapped in \cite{rigby_gaston_2021}. On the figure, we zoom into a small section that contains a strongly varying background and a cloud that we have identified to check how the map is impacted. Despite the complex background emission, the cloud still appears dark with a \fIRB\ equal to 0.05, demonstrating our median filter produces an accurate background model.

\subsection{Verifying \fIRB\ as a tracer of evolutionary stage} 

\noindent As molecular clouds convert a higher fraction of gas into stars, their instantaneous star formation efficiency (SFE) increases along with their dust temperature. Therefore, we use instantaneous SFEs and dust temperatures to test how well \fIRB\ traces evolution by checking if all three quantities correlate with each other. To measure dust temperatures, we use \emph{Herschel} dust temperatures from \cite{peretto_initial_2016}. Since the SFE cannot be measured directly as we do not know the exact amount of stars that have formed, we use 70~\micron\ emission within the cloud (or more specifically, the total 70~\micron\ luminosity of Hi-GAL 70~\micron\ compact sources) as a tracer of the stellar mass and divide it by the mass of the cloud instead.  70~\micron\ emission can be used this way due to a chain of results that link the two properties. In summary, the bolometric luminosity of a protostar approximately equals (i.e., is dominated by) its accretion luminosity and its accretion luminosity is proportional to its protostellar mass \citep{evans_spitzer_2009}. Since we do not know its accretion luminosity, we substitute it with its 70~\micron\ luminosity, which was found to be the most reasonable proxy for the accretion luminosity throughout the majority of protostellar evolution \citep{dunham_identifying_2008,ragan_earliest_2012}.

\begin{figure}
	\centering
		\includegraphics[width = \columnwidth]{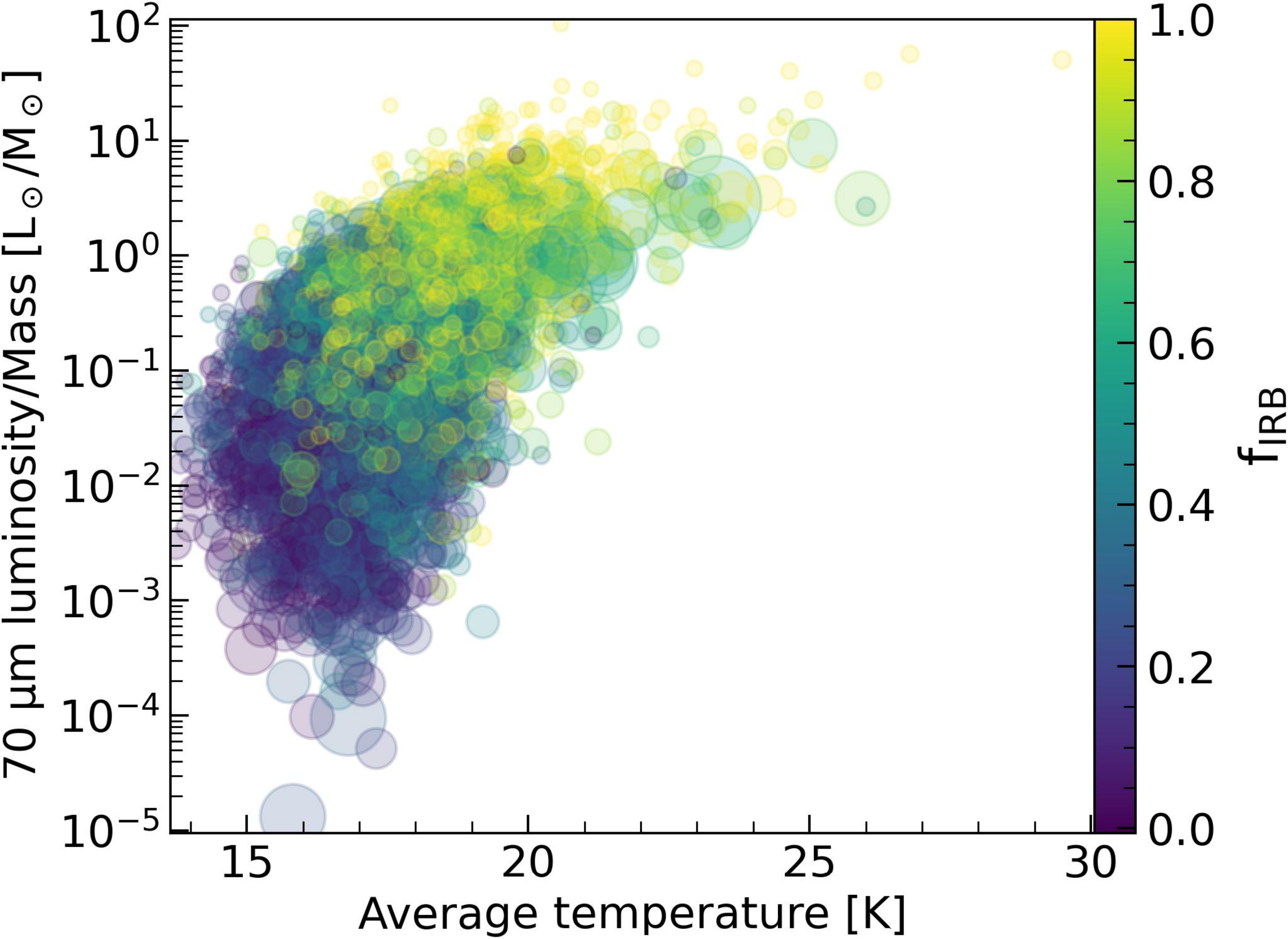}
		\caption{70~\micron\ luminosity to cloud mass against average dust temperature of all clouds with at least one 70~\micron\ source. The colour of the markers indicate their \fIRB\ values and the size of each marker indicates their angular size on the sky.}
	\label{fig:lum_mass_ir}
\end{figure} 
  
 \begin{figure}
	\centering
		\includegraphics[width = \columnwidth]{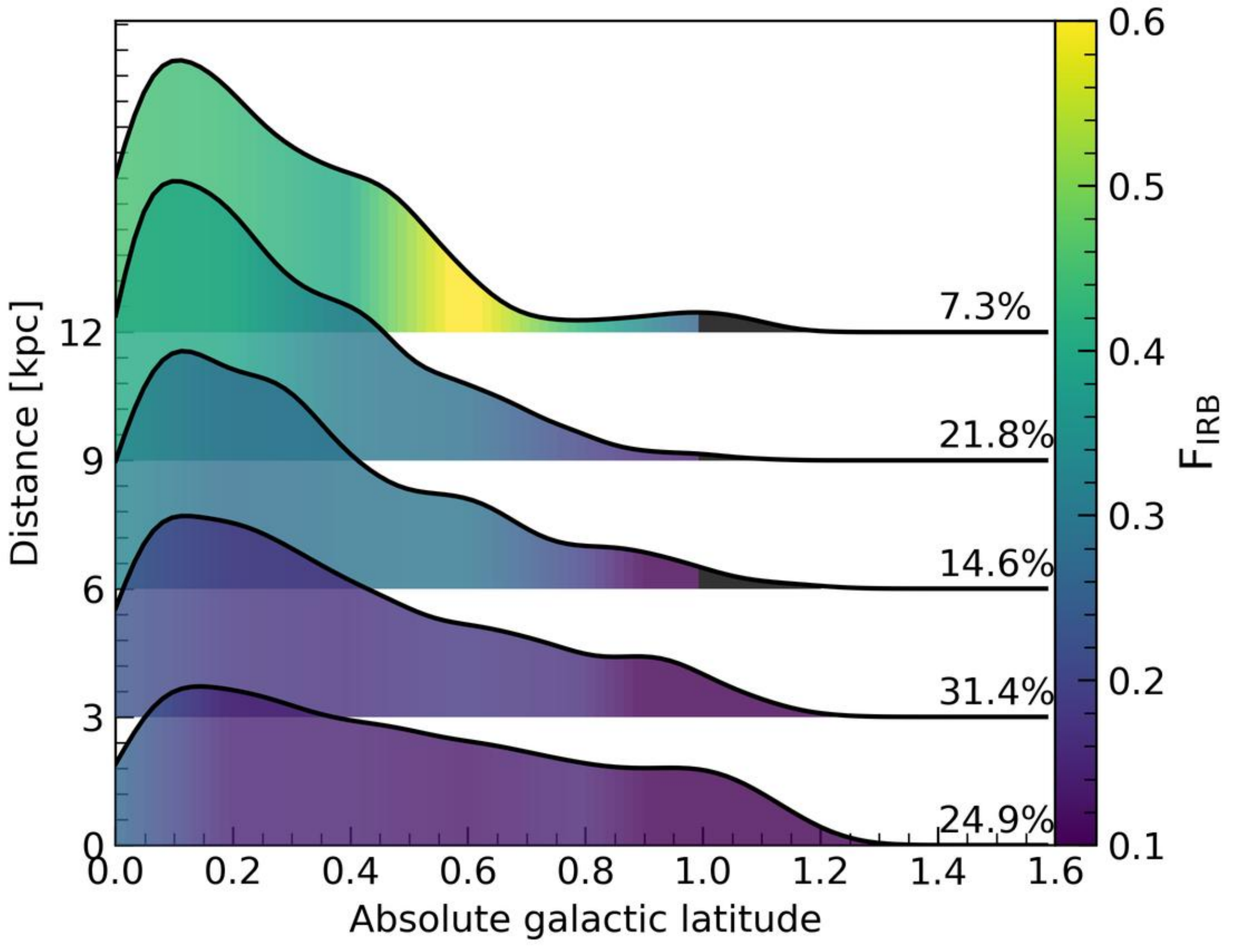}
		\caption{Kernel density estimations of molecular cloud distribution as a function of their Galactic latitude for five distance bins and are coloured by their median \fIRB. Note that the colourmap range is smaller than Fig.~\ref{fig:lum_mass_ir} to enhance contrast. Black indicates a parameter space that contained no clouds. Percentages indicate fraction of clouds contained at each distance.}
	\label{fig:lat-hist-ir}
\end{figure}

\begin{figure*}
	\centering
		\includegraphics[width = \textwidth]{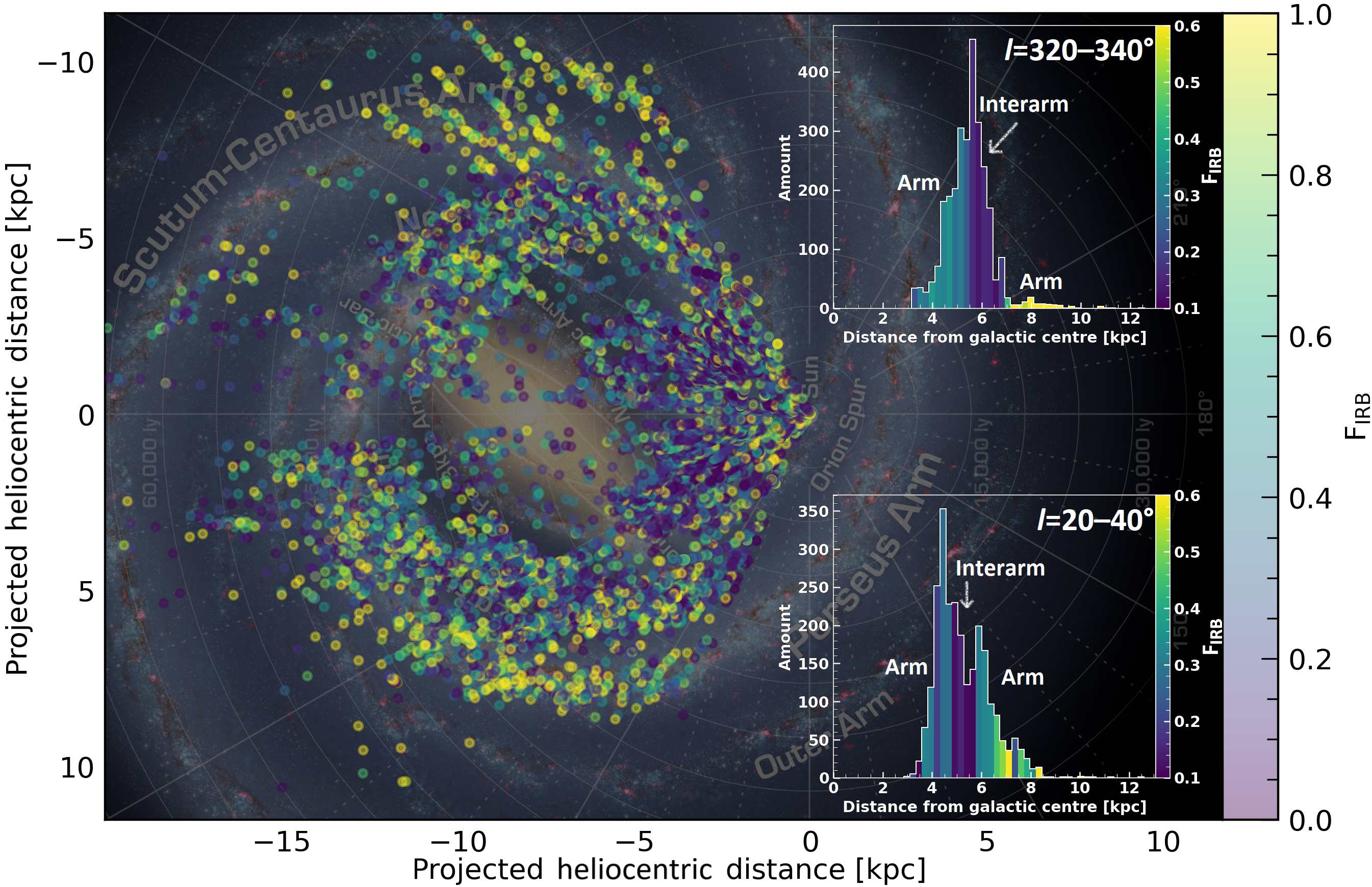}
		\caption{\fIRB\ of molecular clouds over-plotted over the Milky Way. The colour of the markers represent the \fIRB\ of each cloud shown. Inset figures show the cloud number distributions and median \fIRB\ per bin in a narrow longitude range indicated on the inset to emphasise arm to interarm differences. We note here that the very circular distribution of a number of clouds as a function longitude at the tangent point is not real but caused by clouds being assigned the radial velocity at the tangent point if their radial velocity surpassed this value \citep{roman-duval_kinematic_2009}.}
	\label{fig:IR-MW}
\end{figure*} 

\begin{figure*}
	\centering
		\includegraphics[width = \textwidth]{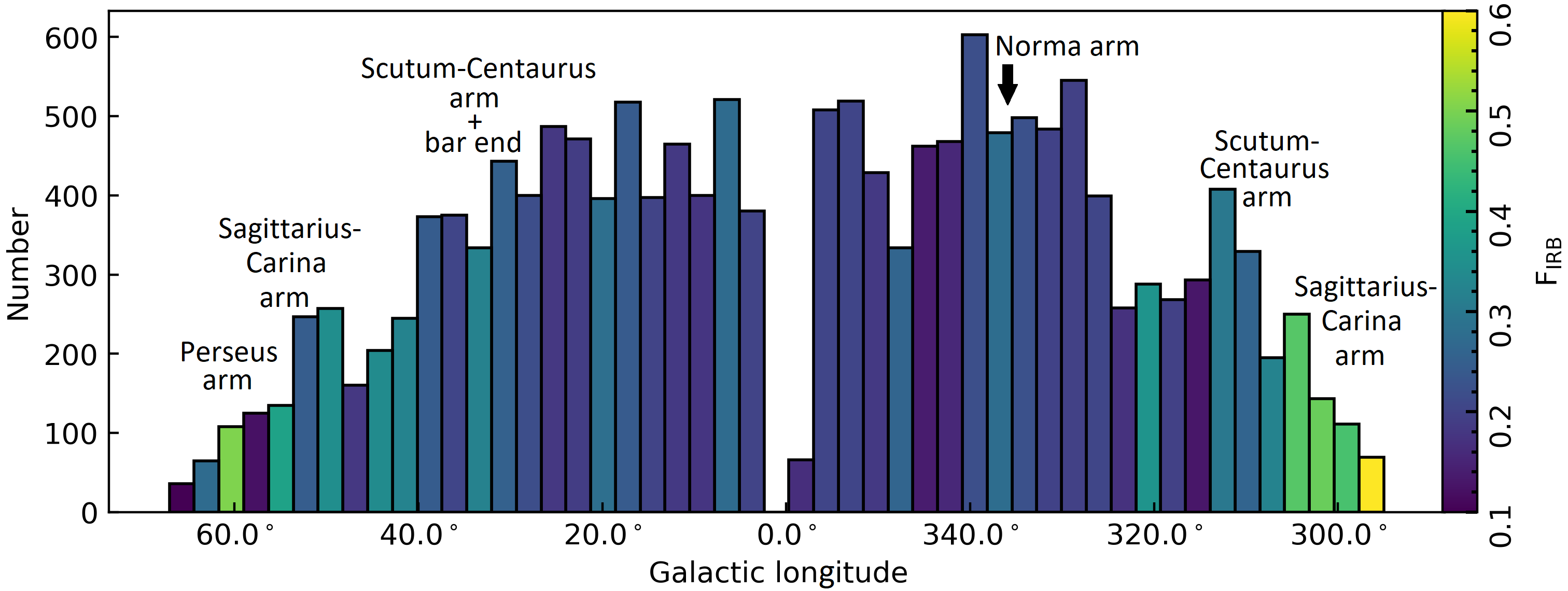}
		\caption{Histogram distributions of \fIRB\ of molecular clouds as a function of galactic longitude. Bins are coloured by their median \fIRB. Expected positions of arms from Fig.~\ref{fig:IR-MW} with narrow longitudes along the line of sight from the Sun are indicated on the plot.}
	\label{fig:IR-long-hist}
\end{figure*}

In Fig.~\ref{fig:lum_mass_ir} we plot the total 70~\micron\ luminosity of Hi-GAL sources over cloud mass against average dust temperature for each cloud and colour the markers by its \fIRB\ value. The marker size also represents the angular area of each molecular cloud to investigate additional factors that could influence the relationship. We note here that only 48\% of clouds had compact 70~\micron\ sources. The figure shows that \fIRB\ correlates with both properties, strongly confirming \fIRB\ ability to trace how evolved the star formation is in molecular clouds. We also see that the 70~\micron\ luminosity over cloud mass values of clouds with larger areas have lower \fIRB\ than smaller clouds. This hints that there are secondary correlations that impact \fIRB. 
	
\subsection{How low background values affect \fIRB} \label{sec:contrast}

\noindent The bright and dark regions at 8~\micron\ are created by two different physical mechanisms. The bright regions are created by the additional photons produced by star formation, which causes them to always appear brighter than the background.
But infrared dark features depend on both the background and foreground emission along the line of sight. Therefore, when the background emission is close to zero, the contrast between the Galactic background and the infrared dark molecular clouds is low, meaning they are more likely to be undetected or be detected with an overestimated \fIRB. Clouds at higher latitudes or at larger distances will be the most affected since 8~\micron\ background will be lower. Using a high column density threshold to catalogue clouds and using a median filtered background to calculate \fIRB\ minimises this bias but we cannot rule out that is has no impact in our catalogue.

To test whether \fIRB\ is impacted by distance or latitude, we plot these two quantities against each other on Fig.~\ref{fig:lat-hist-ir} as a ridge-line plot using a kernel density estimation (KDE). Since clouds are expected to have lower distances at higher Galactic latitudes (due to scale height of the disc) we split the clouds into 3~kpc bins so that each KDE shows the distribution of Galactic latitude between 0--3~kpc, 3--6~kpc etc. We also colour each KDE by the median \fIRB\ binned in 0.2 degree latitude bins and linearly interpolate for a smoother visualisation. Excluding the final distance bin, Fig.~\ref{fig:lat-hist-ir} reveals that \fIRB\ \emph{decreases} as the latitude increases (i.e., as the background emission decreases), indicating that the decreasing background does not make the clouds appear more infrared bright. It implies that the star formation within clouds is slightly less evolved at higher latitudes \citep{planck_collaboration_planck_2011,collaboration_planck_2016,xu_scarcity_2024}. The figure also shows that \fIRB\ does increases with distance.

To investigate if distance bias is responsible for increasing \fIRB, we plot the position of each cloud projected on top of the Milky Way in heliocentric coordinates and label each cloud marker by its \fIRB\ on Fig.~\ref{fig:IR-MW}. While \fIRB\ increases with distance, the spatial information indicates that clouds with projected distances forming arc-like structures, which likely reflect spiral arm features, also have higher \fIRB. It also shows that in between the arm-like features (clouds in `interarm' areas) \fIRB\ is lower. For example Sagittarius-Carina arm at $\ell\sim\ang{300;;}$ reveals that \fIRB\ is high for both nearby and farther distances (on Fig.~\ref{fig:IR-MW}). To quantify this trend, we over-plot two histogram insets on Fig.~\ref{fig:IR-MW} in Galactocentric coordinates instead between $\ell$=20--\ang{40;;} and $\ell$=320--\ang{340;;} and colour the distance bins by the median \fIRB. These longitude ranges include both arm and interarm environments and are able to show that \fIRB\ increases for clouds found on expected arm locations while decreasing in-between them. Finally, we plot the change in \fIRB\ as a function of Galactic longitude as a histogram similar to before on Fig.~\ref{fig:IR-long-hist}. The figure shows when there is a long stretch of spiral arm or a long stretch of interarm on Fig.~\ref{fig:IR-MW} over a narrow Galactic longitude range, \fIRB\ is higher or lower, respectively. Again, the increase in \fIRB\ is especially true for the Sagittarius-Carina arm at $\ell\sim\ang{300;;}$.

Altogether, these figures reveal that there is an increase in \fIRB\ as a function of distance, but we are unable to determine the exact contribution that observational bias has on this trend because the occurrence of arm environments also increases with distance. We can say however that there are physical causes that must play a role in increasing \fIRB\ with distance, considering we see environmental correlations, and considering that the weaker contrast present at higher latitudes does not cause \fIRB\ to increase. For further discussion on the impact distance bias might have on the following results, see Appendix \ref{sec:checks}.

\section{Cloud properties as a function of their \fIRB} \label{sec:cloud_props}

\subsection{Measuring cloud properties and their trends} \label{sec:fIRB_props}
\noindent For each cloud, we first calculate their distance-independent properties such as their projected angular area, their peak and mean column densities ($N_\text{peak}$, $\overline{N}$), their peak and mean dust temperatures ($T_\text{peak}$, $\overline{T}$) and their aspect ratio, $AR$ using their moments. Using the assigned distance, we then calculated the total cloud mass and their effective radius, $r_\text{eff}$, given by:

\begin{equation}
\centering
r_\text{eff} = \left(\cfrac{A_\text{cloud}}{\pi}\right)^{1/2}
\label{eq:reff}
\end{equation}

\noindent where $A_\text{cloud}$ is the cloud area. Using the effective radius, we also estimate the mean spherical number density of each cloud. We then calculate their mass using the same molecular weight used in \cite{peretto_initial_2016} of 2.8. After assigning cloud properties, we find a small number of clouds had no assigned number, or infinity for properties such as their mass, or their column density. These are caused by unassigned, or infinite values within the \textit{Herschel} maps when saturation occurred in the observation itself. This removes an additional 405 clouds from our sample. Our final working sample is therefore 12261.

\begin{figure}
	\centering
		\includegraphics[width = \columnwidth]{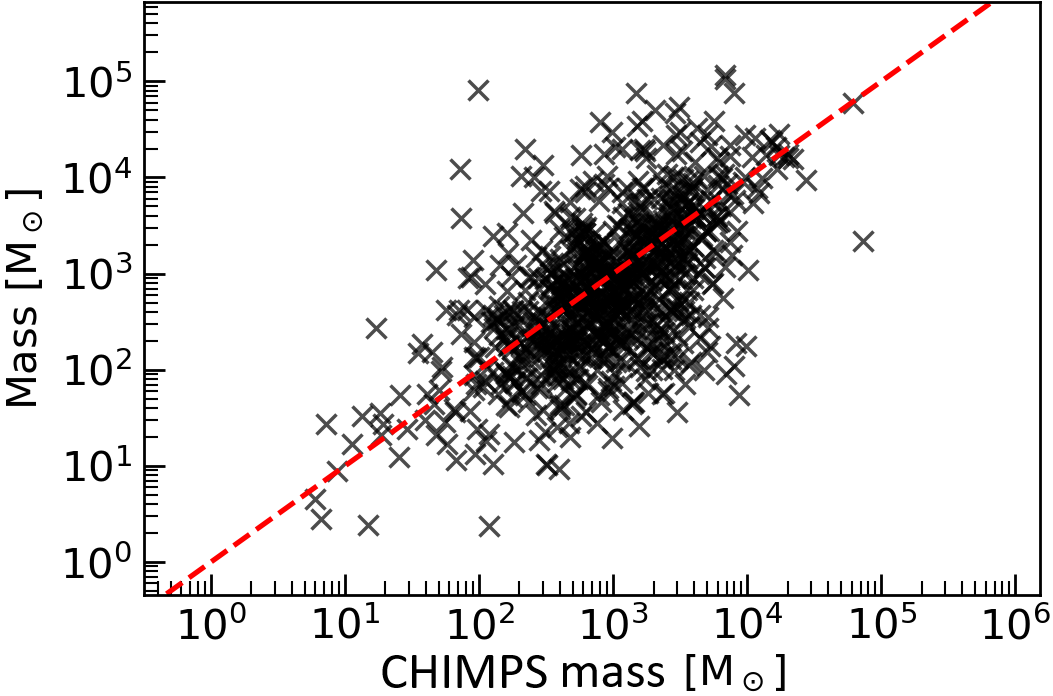}
		\caption{Mass of our catalogue against matched clouds from the CHIMPS catalogue \citep{rigby_chimps_2019}. Dashed red line shows the one-to-one line.}
	\label{fig:mass-vs-chimps}
\end{figure} 

\begin{figure}
	\centering
		\includegraphics[width = \columnwidth]{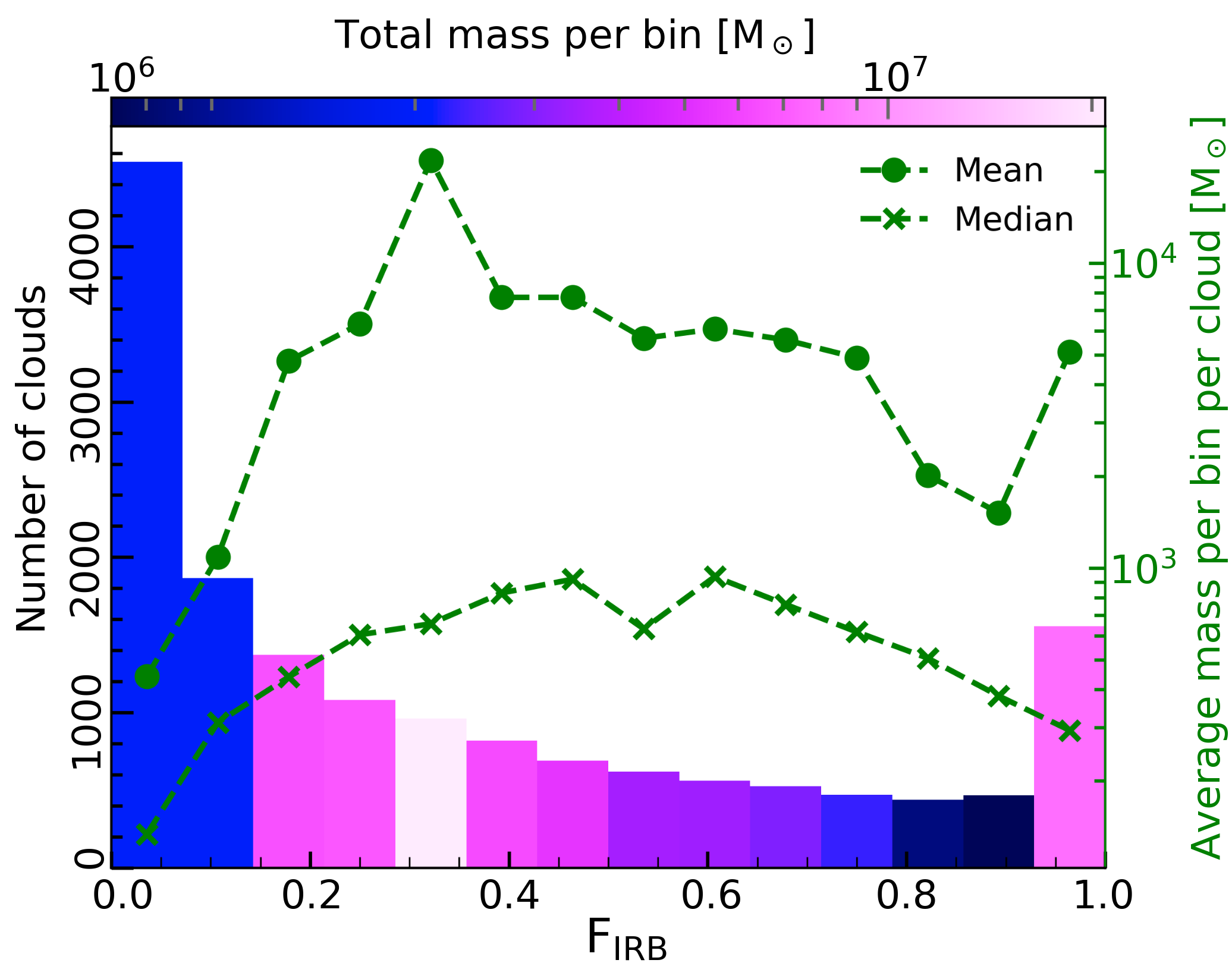}
		\caption{Distribution of \fIRB\ values with bins coloured by the total cloud mass within that bin. Over-plotted in green lines are the average mass of clouds per bin. The circle markers show the mean values per bin. Cross markers show the median mass per bin.}
	\label{fig:hist-ir}
\end{figure} 

To confirm that we are tracing physically realistic clouds, we compare the masses we derive to the CHIMPS catalogue \citep{rigby_chimps_2019}. CHIMPS is a molecular cloud catalogue that identified clouds using $\ang{;;15}$ resolution $^{13}$CO (3--2) data between $\ang{27.8} \leq \ell \leq \ang{46.2}$ and $|b|\leq 0.5$ using the {\tt FellWalker} \citep{berry_fellwalkerclump_2015} algorithm (see \citealt{rigby_chimps_2019} for more details). 
CHIMPS represents the ideal comparison set since $^{13}$CO (3--2) traces higher column density structures, which \cite{rigby_chimps_2019} results in them identifying clouds with sizes and densities intermediate between small- and large-scale molecular clouds. We cross-match the two catalogues only using the CHIMPS clouds that had a reliable flag. To cross match, we use the central coordinates of both with a radii of $\ang{;;21}$ each, matching the smoothing scale used to generate our catalogue in Sect. \ref{sec:cloud_id}. We find 1115 out of 3664 CHIMPS clouds overlap with our catalogue and we plot their mass on Fig.~\ref{fig:mass-vs-chimps} using the distances derived in Sect. \ref{sec:dist}. We find the two have good correspondence with the one-to-one line. Using Kendall's Tau, we find they have a correlation of 0.40 with a p-value of 8.8\e$^{-69}$, therefore we are confident our catalogue is representative of molecular clouds at higher column density thresholds.

Next, we explore the distribution of \fIRB\ and how it relates to the cloud properties. Even though clouds are dynamic and likely interchange mass via collisions and mergers during their lifetime \citep{jeffreson_clouds_2023}, for simplicity, in this section we assume that \fIRB\ linearly traces the lifetime of star-forming molecular clouds. We also assume our catalogue contains enough independent star-forming regions so that each phase is statistically well sampled. Under these assumptions, we expect an even number of bright and dark clouds. But when plotting the histogram distribution of \fIRB\ on Fig.~\ref{fig:hist-ir}, we find more clouds are infrared dark than bright. For example, 71\% of clouds possess \fIRB$<0.5$ and the lowest bin (\fIRB\ between 0--0.07) represents 27\% of the catalogue, These results imply clouds spend more time starless. We also see that many clouds that have \fIRB\ values in the lowest \fIRB\ bin contain average column densities close to 1\e$^{22}$~cm$^{-2}$. These clouds represent the highest column density peaks from a sample of more extended, lower density clouds that our high column density cloud identification threshold excludes. Lower density clouds have longer free-fall times and so remain starless for longer, providing an even stronger argument for a longer starless period.

Similar results using \fIRB\ were also found in \cite{rigby_gaston_2021} who found the 75\% of clumps have \fIRB<0.663. In addition, \cite{battersby_lifetimes_2017} also show that the infrared dark stage contains more gas and by using nearby maser measurements, they were able to assign lifetimes to these stages to confirm that the lifetime of infrared bright gas (i.e., gas containing star formation) is shorter than infrared dark gas. Observations that also characterise SFE per free-fall time ($\varepsilon_\text{ff}$) indicate $\varepsilon_\text{ff}$ increases with as clouds age \citep{lee_formation_2016}. With \fIRB\ correlating with SFE in Fig.~\ref{fig:lum_mass_ir}, our results are consistent with a longer quiescent phase.

We next investigate how the mass is distributed within these bins, including the total mass per bin in Fig.~\ref{fig:hist-ir}, the mean mass per bin and the median mass per bin. Both the total and mean mass per bin peak when \fIRB\ is at intermediate values at around \fIRB$\sim$0.35, not at 0, and the median mass per bin peaks at \fIRB$\sim$0.5. The difference between the mean, median and total indicates that a small number of clouds have a much more rapid growth phase that skews the mean (and therefore increases the total mas per bin) to higher mass values. If clouds are isolated objects, we expect them to have more mass when their star formation is unevolved, more so considering that the majority of clouds in our sample are at starless stages (a quarter of the entire sample has \fIRB<0.07). Then as stars evolve, stellar feedback is predicted to remove the gas. Instead, these results show that on average, all clouds gain molecular gas as star formation progresses rather than losing mass, at least until 50\% of their area has star formation. The decrease in number distribution of clouds on Fig.~\ref{fig:hist-ir} also implies that the rate of mass gain must accelerate in time. In summary, these results show molecular clouds are not isolated but instead grow in mass over time while forming stars.

\begin{figure*}
	\centering
		\includegraphics[width = \textwidth]{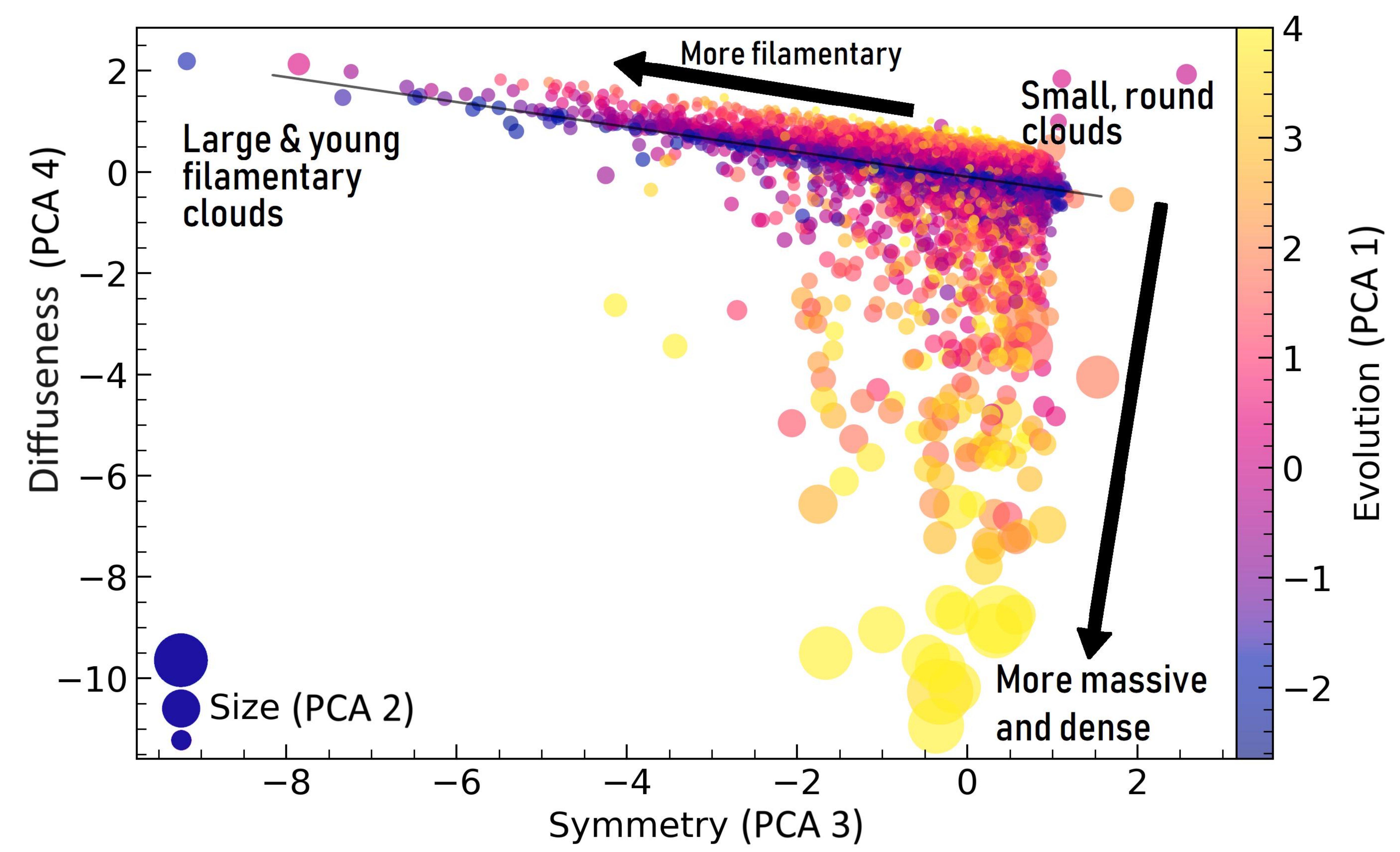}
		\caption{Score plot of the cloud catalogue using their principal components. The diffuseness axis (PC4) is represented using the y axis, the symmetry axis (PC3) is represented using the x axis, the marker colours represent the evolution axis (PC1) and the size of the markers shows the size of the clouds (PC2). This black line illustrates the evolutionary minimum. Annotations indicate the cloud properties present at those plot locations.}
	\label{fig:pca4}
\end{figure*}

\subsection{Exploring data trends using PCA analysis} \label{sec:pca}
Results presented previously in this Section and Section \ref{sec:firb} indicate there are secondary correlations impacting the cloud properties. For example, Fig.~\ref{fig:lum_mass_ir} has a secondary correlation relating to the cloud size. To explore the correlations between the cloud properties, and to optimise how trends are displayed, we perform Principal Component Analysis (PCA) on the cloud properties. PCA is a multivariate method that linearly combines a set of variables along successive, orthogonal axes to maximise the variance of the data set. PCA implicitly assumes data linearity and that the components are orthogonal, both of which are needed for simple, solvable PCA solutions. It also assumes that data variance represents information, so variables with larger variances are more important and necessary to explain the data. By projecting the data along these new axes -- their principal components -- we are maximising the information we can access (i.e., where a small increase or decrease in the variables result in the largest change in the cloud properties). This allows us to reduce the dimensionality of the data by removing principal components we know contain the least variance, which simplifies data trends while retaining the most amount of information possible. It also projects the data along the axes that maximise the variance, which maximises the visibility of any trends present.

\begin{table}
\resizebox{\columnwidth}{!}{
    \begin{tabular}{l|ccccc} 
    Variable & PC1 & PC2 & PC3 & PC4 & PC5 \\
     & (Evolution) & (Size) & (Symmetry) & (Diffuseness) &  \\
    \hline
    Peak column density & 0.29 & 0.55 & -0.03 & -0.78 & -0.10  \\
    Cloud mass & 0.08 & 0.57 & 0.71 & 0.40 & 0.04 \\
    Average temperature & 0.68 & -0.10 & -0.09 & 0.10 & 0.71 \\
    IR brightness fraction & 0.66 & -0.15 & -0.04 & 0.23 & -0.69 \\
    Aspect ratio & -0.06 & 0.59 & -0.70 & 0.41 & -0.00 \\
    \hline
    Explained variance & 30\% & 21\% & 20\% & 18\% & 11\%
    \end{tabular}
    }
    \caption{Table showing the eigenvectors of all five principle components (and their physical interpretation) for each variable. Values close to 0 indicate little contribution from the variable to the principle component, while values close 1 or -1 indicate strong contribution from the variable.}
    \label{tab:vectors}
\end{table}

The principal components are calculated by scaling the data, finding the eigenvectors of the data covariance matrix, and ordering them according to their eigenvalues, from largest to smallest. The data are then converted to their principal component scores by projecting the data to this eigenbasis. The eigenvectors, $\vect{v}_i$, therefore point in the direction of the principal components while the absolute value of eigenvalues |$\lambda_i$|, -- which we normalise so that their sum equals one -- explains the variance that component represents.

Before we perform PCA, we first eliminate variables that are directly derived from a more fundamental variable to ensure data orthogonality, such as the number density, which is derived from the mass and the physical cloud area. The mass also depends on the distance and the physical area so there will be no new information gained by including the last two variables. Additionally, both the average and peak temperature, and the average and peak column density are highly correlated parameters, information we know and expect. Including them paired together will skew the PCA towards them. After testing the peak and average values separately using PCA, we find that the average temperature, and the peak column density produced stronger components. Altogether, we performed PCA using \fIRB, the peak column density, the average temperature, the cloud mass and the aspect ratio of the clouds and tabulate the eigenvectors (i.e., the principle directions) and their absolute normalised eigenvalues (i.e., their explained variance) in Table \ref{tab:vectors}.

Our first check is to find and remove principal components that only explain a small fraction of the data variance to simplify the interpretation and visualisation of the data. We find that there is no single component that explains the majority ($\sim$90\%) of the variance, which would greatly simplify our interpretation. Instead, we find that the almost all the data variance can be explained with the first four principal components, cumulatively representing 89\% of the data variance and show this in Table \ref{tab:vectors}. As a result, we remove component five and project the data without it. With the four remaining components, we interpret the correlations shown in Table \ref{tab:vectors}. Table \ref{tab:vectors} shows:
\begin{itemize}
    \item The first principle component (PC1) is dominated by \fIRB\ and the average temperature, with a small contribution from the peak column density. Since \fIRB\ and the temperature trace evolution, we call PC1 the `evolution' axis; 
    \item PC2 traces the mass, peak column density and the aspect ratio of the clouds, all of which are size related parameters. Objects with more mass tend to reach higher column densities and so we call this axis the `size' axis; 
    \item PC3 traces a strong anti-correlation between mass and aspect ratio. This component likely shows that spherical clouds contain more mass than filamentary clouds when both have the same column density (i.e., spherical clouds have a larger area). As a result, we call this axis the `symmetry' axis; 
    \item PC4 shows that an increase in the mass and the aspect ratio results in a decrease in peak column density. The physical interpretation we infer is rounder objects require less mass to reach higher peak column densities. Therefore, we call this axis the `diffuseness' axis so that negative values indicate a cloud that is more centrally concentrated. 
\end{itemize}
\noindent Potentially, PC3 and PC4 both indicate that spherical clouds are more efficient at concentrating mass, which was also found in \cite{rigby_nika_2018}.

\begin{figure}
	\centering
		\includegraphics[width = \columnwidth]{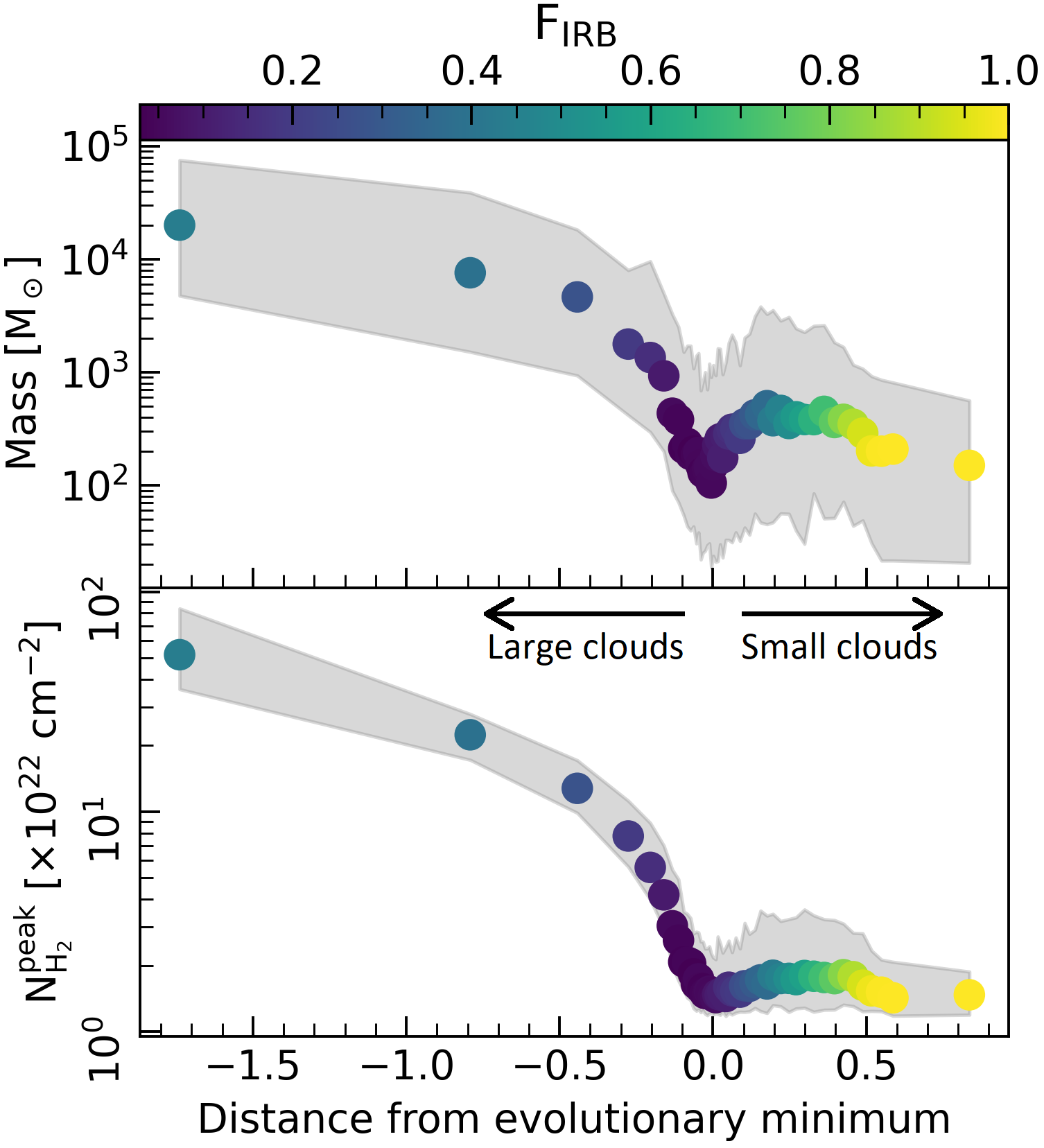}
		\caption{Median cloud properties as a function of evolution. Median calculated using equal number of clouds binned as a function of distance  perpendicular from the evolutionary minimum represented as the black line on Fig.~\ref{fig:pca4}. Therefore negative values show the change in cloud properties as a function of evolution for clouds that fall below the evolutionary minimum and positive values show the same but for clouds above. \textit{Top panel} are median mass values and \textit{bottom panel} are median peak column density values. The marker colours show the median \fIRB\ and the grey-shaded region show the 16-84\% sigma spread in the mass and peak column density.}
	\label{fig:yinc}
\end{figure}

To explore how the cloud properties relate to the four principal axes, we plot the cloud scores projected along all four principal axes on Fig.~\ref{fig:pca4}. We plot the diffuseness against the symmetry axis and colour the markers by their evolution and change the marker size to show the size axis. The graph shows that clouds have a distinct location as a function of their morphology and evolutionary state. We find the youngest, and least evolved clouds on a narrow horizontal band. Their position in respect to the symmetry and diffuseness axis indicate that the youngest clouds have a range of aspect ratios, though clouds with high aspect ratios (i.e., more filametrary) have lower peak column densities but contain more mass.

The most interesting trend on Fig.~\ref{fig:pca4} however is how the symmetry and diffuseness change as the clouds evolve. Firstly, we see that clouds appear to become more evolved with different properties. The clouds that evolve down and left on the figure get larger, become more centrally concentrated (increase their peak column density) but become less symmetrical. The clouds that evolve up and right on the figure only increase their mass and peak column density slightly as they evolve and the rightward direction means they also become slightly more symmetric.

To better understand this evolutionary trend and how it relates to the physical cloud properties, we calculate the median mass, peak column density and \fIRB\ in parallel strips (i.e., bins) away from the evolutionary minimum, which is indicated using the thin solid black line on Fig.~\ref{fig:pca4}. Each parallel strip contains an equal number of clouds. We plot these averages on Fig.~\ref{fig:yinc} where the y axes on the top and bottom panels show the median mass and peak column density respectively while the x axis shows distance away from the evolutionary minimum. A value of zero on the x axis of Fig.~\ref{fig:yinc} therefore represents the centre of the evolutionary minimum, negative values show clouds that evolve downwards on Fig.~\ref{fig:pca4}, and positive values show clouds upwards. The marker colours show the median \fIRB\ values.

Figure \ref{fig:yinc} shows that the cloud properties bifurcate. While both directions show clouds grow as star formation occurs, the clouds that evolve vertically upwards (positive values on the x axis of Fig.~\ref{fig:yinc}) gain up to $\sim$4 times their starting mass, while their peak column densities only reach a factor of 1.3 times their starting value on average as \fIRB\ increase. After \fIRB$>$0.5, the total mass decreases but their peak column density remains constant until the cloud becomes \fIRB$\simeq$0.9 where their peak column density finally begins to decrease. Potentially, this shows that when dense gas forms, it is more resilient to being removed, while the more diffuse mass can be removed after the clouds become star-forming \citep{watkins_feedback_2019}. The majority of clouds (which we refer to as `small clouds') follow these average evolutionary properties, representing $\sim$80\% of the catalogue. Indeed, the mass resembles the median averaged cloud mass per \fIRB\ bin on Fig.~\ref{fig:hist-ir}.

This is the simple, standard scenario that might be expected for the star forming clouds: clouds collapse due to gravity to form stars, and then are destroyed by the feedback the stars produce. Our results add that the clouds continuously accretes material while forming stars, and this halts only after 50\% of the cloud area is star forming. However a more isolated view of star formation is especially unsuited in 20\% of cases. These clouds gain significantly more mass (a factor of $\sim$150 times more) and reach much higher peak column densities (becoming 40 times more dense) as \fIRB\ increases. In fact, Fig.~\ref{fig:yinc} reveals that in the beginning, the average \fIRB\ value \emph{decreases} slightly for the high-mass clouds. This suggests that initially, mass rapidly accretes onto these clouds at a rate that outpaces the initial star formation occurring within the clouds. While weaker, the same trend can be seen in Fig.~\ref{fig:lum_mass_ir}. The figure shows that clouds with a larger angular size have lower \fIRB\ than the smaller clouds. This shows how PCA is able to disentangle subtle trends by maximising the variance.

Because these clouds are significantly larger (5 times more massive on average), they represent 55\% of the mass contained in this survey. Assuming that all clouds have the same SFE, it indicates that half of all star formation in our sample occurs within this small population of molecular clouds. As a result, we refer to these clouds as `large clouds'. \cite{miville-deschenes_physical_2017} and \cite{urquhart_atlasgal_2018} also show that most of the star formation within the Milky Way is found in a small number of massive star-forming complexes.

\begin{figure}
	\centering
		\includegraphics[width = \columnwidth]{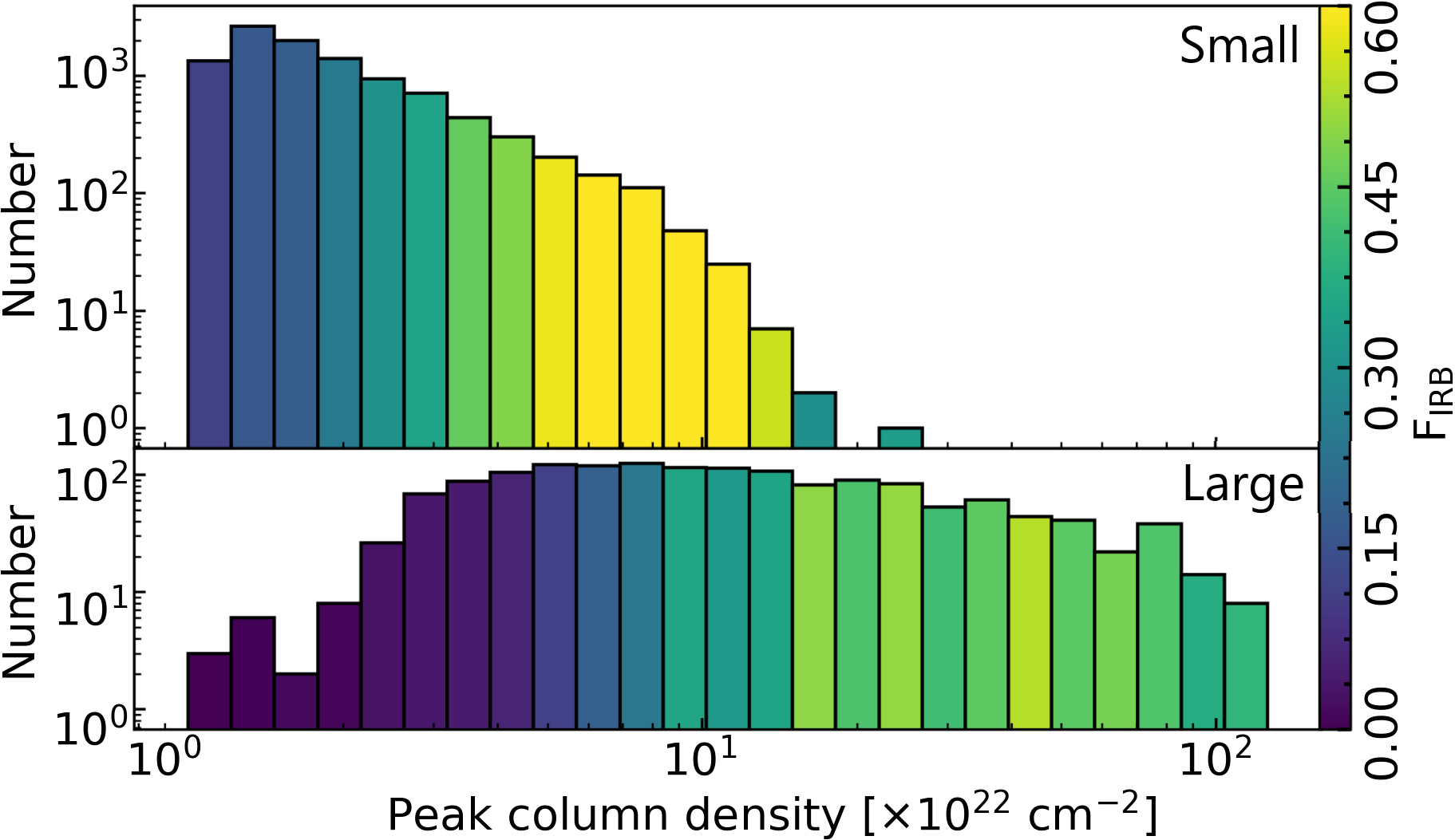}
		\caption{ Histogram distribution of peak column density values for \textit{large clouds} and \textit{small clouds}. Bins coloured by the median \fIRB\ per bin for the two cloud types revealed in Fig.~\ref{fig:pca4}. \textit{Top panel} are small clouds (clouds that evolve above the line on Fig.~\ref{fig:pca4}) and \textit{bottom panel} are large clouds (clouds that evolved below the line instead). Size of the y axis on both panels use the same scale to better compare the distributions.}
	\label{fig:hist-ir-pCD-sp}
\end{figure}

\begin{figure}
	\centering
		\includegraphics[width = \columnwidth]{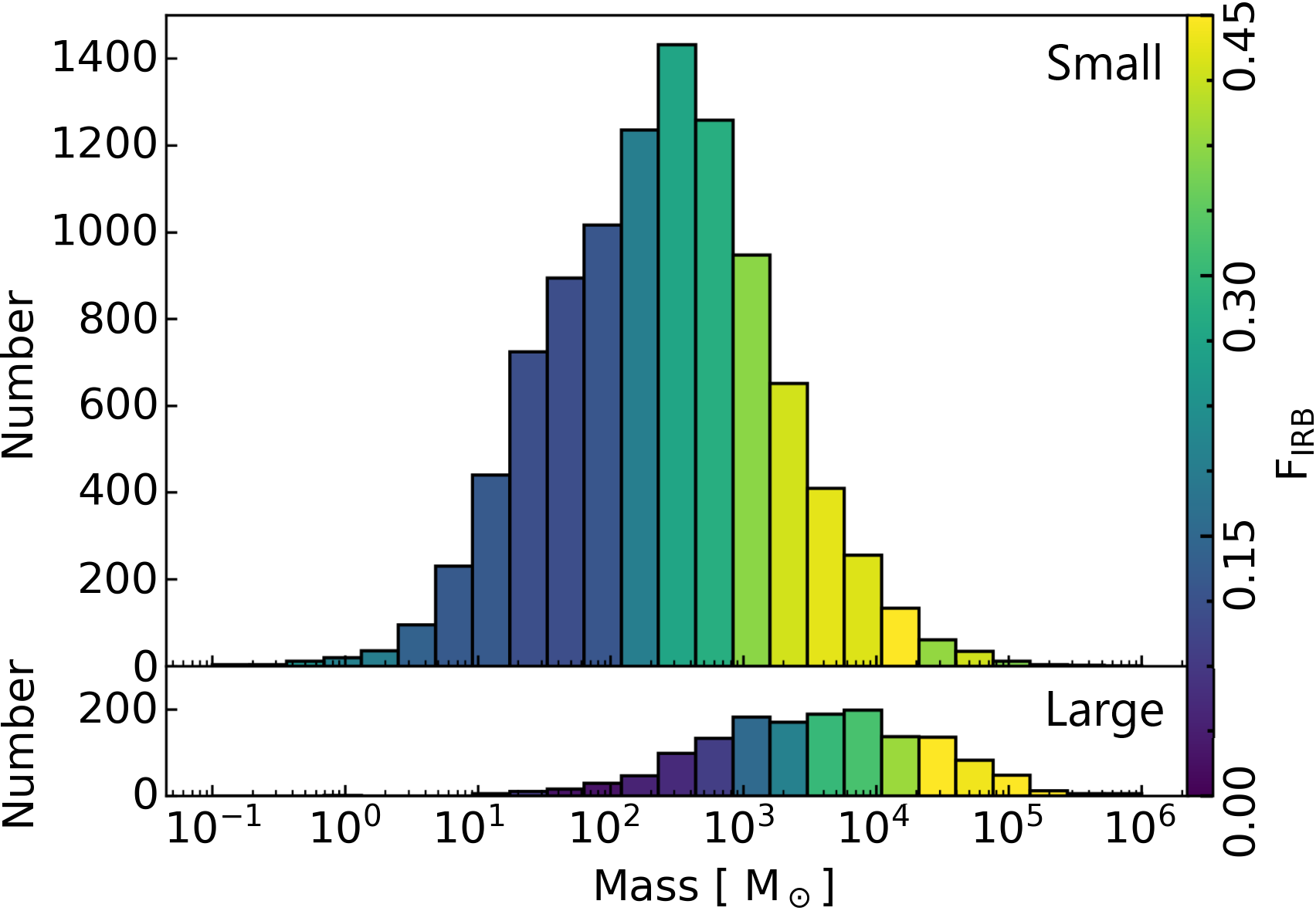}
		\caption{Same as Fig.~\ref{fig:hist-ir-pCD-sp} but for mass.}
	\label{fig:hist-ir-mass-sp}
\end{figure}

While we see the cloud growth on Fig.~\ref{fig:hist-ir} and discuss it in Sect. \ref{sec:cloud_props}, to confirm that there is two different amounts that clouds grow by (and is not an artefact of the PCA projection), we plot the histogram distributions for peak column density and mass, and colour the bins by their median \fIRB\ per bin for the two cloud types in Fig.~\ref{fig:hist-ir-pCD-sp}, Fig.~\ref{fig:hist-ir-mass-sp}. The two plots show that there are distinct differences in both the distribution of values and the median \fIRB\ for the two cloud types. For example, Fig.~\ref{fig:hist-ir-pCD-sp} shows that the bigger clouds have much lower \fIRB\ when peak column densities are small and have a much larger tail for high values. We end this section by noting that we investigate if our previous analysis impacts the PCA results in Appendix \ref{sec:checks}. In particular, we check if the bifurcation could be caused by the cloud merging algorithm used in Sec.~\ref{sec:cloud_id}, if any of our results are impacted by the distance assigned, and if the potential distance bias (discussed in Sect. \ref{sec:contrast}) impacts the PCA. Our investigations show that our analysis does not impact the PCA results.

\begin{figure*}
	\centering
		\includegraphics[width = \textwidth]{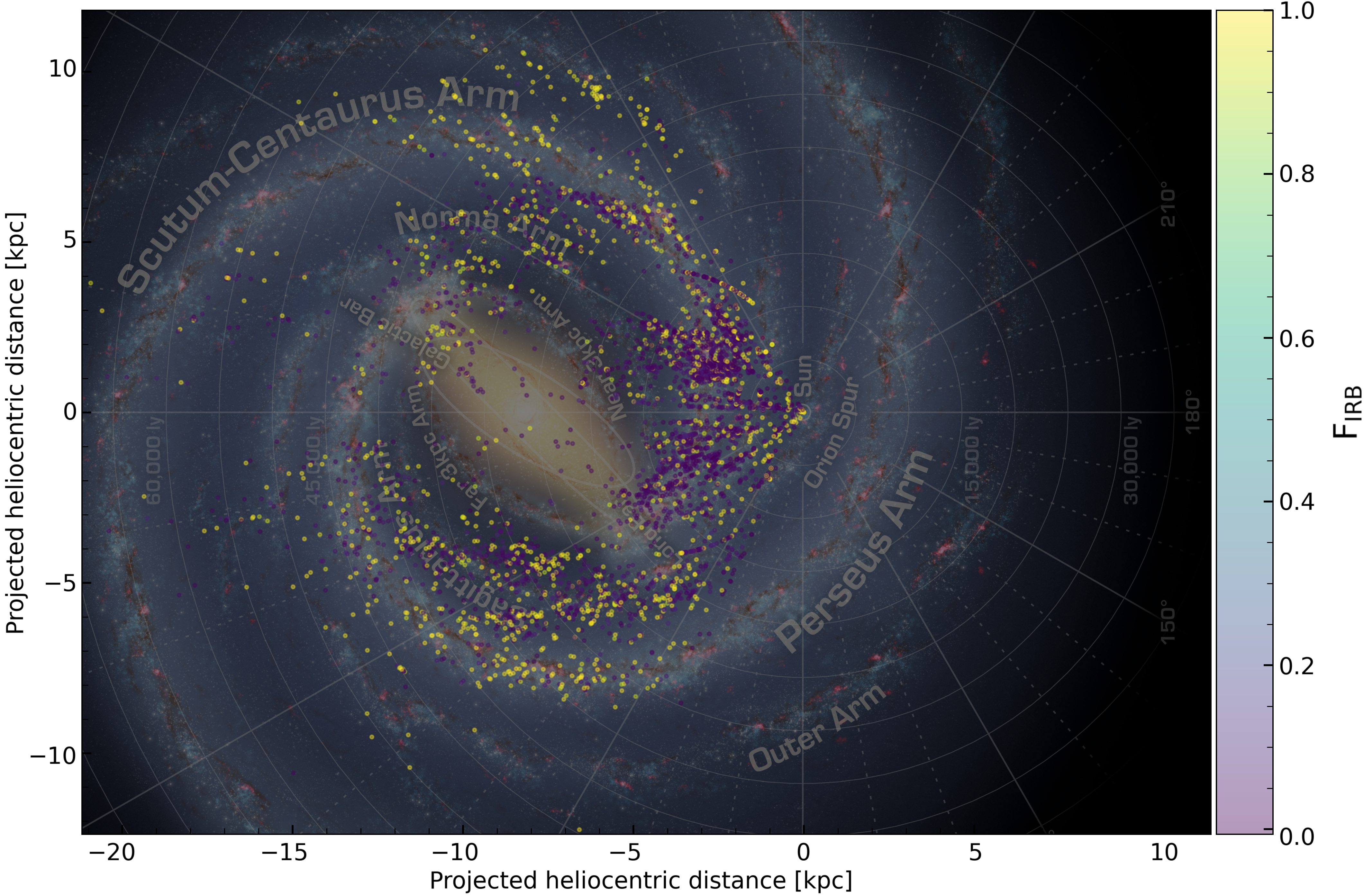}
		\caption{\fIRB\ of molecular clouds over-plotted over the Milky Way only for clouds with high (>0.90) and low (<0.1) \fIRB values. Blue markers show low \fIRB\ clouds while yellow show the high \fIRB\ clouds.}
	\label{fig:IR-MW-evo}
\end{figure*}

\subsection{Environmental trends} \label{sec:env}
Here, we explore potential trends in cloud properties vs. environmental features and discuss the implications of our results. We note that since the location of spiral arms, or even the number of arms is are still a matter of debate, and kinematic distances can be problematic \citep{reid_accuracy_2022} the analysis we can perform is limited. Therefore, we do not compare against any particular model other than the Milky Way visualisation as an aid to the discussion and use the large scale structure of our cloud catalogue projected on a face-down view to inform whether features are likely associated with arm or interarm environments.

In Fig.~\ref{fig:IR-MW} we find that \fIRB\ increases in clouds grouped together into arc-like features, which when over-plotted on a face-down visualisation of the Milky Way, show alignment with spiral arm features. We also show that between these arcs, \fIRB\ of clouds decreases in visual agreement to more interarm environments. This is more easily seen on Fig.~\ref{fig:IR-MW-evo} where we plot the spatial distribution of our clouds that have \fIRB<0.1 or \fIRB>0.90. We can see how the most evolved clouds cluster together into thin arcs and, more often than not, between these arcs are the less evolved clouds. If we assume that all clouds goes through the entire range of \fIRB\ values during their lifetimes, Fig.~\ref{fig:IR-MW-evo} shows that the relative duration of the different phases changes as a function of Galactic location. This likely represents a difference in cloud mass growth, implying the larger scale structure influences how molecular clouds evolve.

In Appendix \ref{sec:mw-birdseye} we further investigate the distribution of the two cloud types between the arms and interarm regions by plotting their top-down positions on Fig.~\ref{fig:IR-MW-large} and Fig.~\ref{fig:IR-MW-small}. These figures show that the large cloud distribution form tight extended arc-like structures that congregate to the expected positions of the spiral arms illustrated on the visualisation of the Milky Way. Furthermore, the face-down distribution shows that the arcs are separated by gaps. 
The lower-mass clouds do not have as strongly delineated arc features, appearing everywhere, though they still cluster into arc-like features and we visually see small clouds before an arm-like feature have lower \fIRB\ values, and larger \fIRB\ on and after it. 
The large clouds therefore seem to preferentially formed in spiral arms. This could be due to spiral arms enhancing the conditions necessary for their formation. For example, an increase in surface density of molecular clouds was also detected in \cite{miville-deschenes_physical_2017} using CO. However, we cannot yet rule out the large clouds being a random sampling of the small cloud distribution given their more limited numbers.

\section{Discussion} \label{sec:discuss}
\subsection{Molecular cloud growth during star formation.}
From Fig.~\ref{fig:hist-ir-mass-sp} it is clear that both the small and large clouds have greater masses when \fIRB\ is higher. This suggests that both populations of clouds are growing in mass \emph{at the same time} as they are undergoing star formation. Let us now consider how the cloud mass and star formation evolve for both types of clouds.

We find that small clouds grow by four times their original mass as their star formation evolves. While the exact mechanism for this growth is unclear, it can be explained easily by cloud collisions or by clouds accreting the surrounding lower density H$_2$. The time scale for cloud-cloud collisions is simulated to be 8--10~Myr in spiral galaxies \citep{dobbs_frequency_2015}. If we estimate that small clouds recycle their mass at least every 40~Myr from the feedback injection timescale from clustered supernovae \citep{orr_bursting_2022} they would be expected to collide $\sim$4 times over their lifetime, matching the expected grow rate we see. We can also explain the growth by assuming the small clouds accrete the surrounding gas. If it accretes the surrounding 1\e$^{21}$~cm$^{-2}$ column density gas within a 3~pc volume, we can achieve mass growth rates matching our results. However, what is more difficult to explain is why they only gain four times their mass. Potentially, if only part of the structure is gravitational bound and able to form stars, they will not effectively convert their gas into stars. Moreover, their diffuse nature may make them more easily disrupted by external forces such as supernovae \citep{feng_evolution_2024} or tidal forces \citep{ramirez-galeano_why_2022}. Thus the lowest mass part of the sample may never evolve into massive clouds.

Similar growth arguments are harder to apply to large clouds. One scenario is that the large clouds form from a single larger scale instability. Therefore, we might be witnessing clouds condensing at a higher mass via gravitational instabilities directly from the H{\sc i} in the high potential environment of the spiral arms \citep{meidt_molecular_2022}. It is from this state they could continue to grow into the large clouds we find in this survey. Large clouds could also represent where H$_2$ formation occurs in a non-equilibrium state \citep{burkhart_molecular_2024}. These authors show molecular clouds can gain over a factor 100 times more mass, while star formation is ongoing. In particular, we note that the mass growth curve shown in their fig.~1 is remarkably similar to the mass growth we see in Fig.~\ref{fig:yinc}. Recent simulations from \citet{dobbs_properties_2011} and \citet{jeffreson_clouds_2023} also show that molecular clouds were found to persist for long periods (up to 90~Myr) despite rapid star formation within them due to fresh accretion of gas and due to mergers with smaller clouds.

Our results favour the latter scenario. We show there is a clear difference in the column density evolution of the two cloud samples shown in Fig.~\ref{fig:hist-ir-pCD-sp}. The small clouds decrease in number as the peak column density increases. This is consistent with single objects undergoing free fall collapse, where the evolutionary timescale is shorter at higher densities. However, the large clouds have a much flatter distribution of peak column densities, with less clouds with low peak column densities, which also have lower \fIRB\ values. 
This suggests that the clouds are gaining fresh mass as regions within them are collapsing. This would replenish the cloud with new infrared dark gas allowing \fIRB\ to remain low while the cloud itself continues to grow to higher masses and column densities. Finally, the absence of high \fIRB\ can be explained by the rapid destruction of the clouds once star formation finally outpaces the mass growth. The high-mass cloud likely consumes and destroys the remain gas, halting mass accretion. As a result, it no longer meets the properties needed to be identified as a large cloud (which is likely responsible for the increase in massive highly evolved star formation in Fig.~\ref{fig:hist-ir}. This also matches the evolutionary path outline in \cite{burkhart_molecular_2024}. They show that cloud growth halts only when enough stars form that they destroy the H$_2$ faster than the mass gain. We therefore expect large clouds would preferentially form in environments where there is a large gas reservoir, crowding of orbits to facilitate merging of clouds, and a deep potential well to encourage large scale collapse. Such an environment may be found in spiral arms. 

Dynamic star forming processes have been predicted before \citep{lee_observational_2016,kobayashi_evolutionary_2017,vazquez-semadeni_global_2019}. For example \cite{lee_observational_2016} found star formation rates increase over time, which may be explained by clouds growing and collapsing while forming stars. These results have strong implications on how we assess the star forming efficiency within a region, and the final number of stars a cloud produces. The mass we use to judge a cloud properties, such as the number of stars it can produce, can be misleading if clouds grow over time alongside their star formation. Furthermore, if
the cloud has recently accreted a significant amount of mass faster than it forms stars, we would predict that it has a low star formation efficiency per free-fall time.

\subsection{How large scale structure impacts cloud properties}

In the previous section (Sect. \ref{sec:env}), we raise the possibility that there may be a change in the evolution of star formation, as traced by \fIRB\, in the vicinity of spiral arms. For example, the histogram insets on Fig.~\ref{fig:IR-MW} hints that the clouds that gather into arc-like structures have more evolved star formation than the rest of the galaxy. The top-down view of the least and most evolved clouds on Fig.~\ref{fig:IR-MW-evo} also hints at this relationship assuming that the long arc-like structures that the projected cloud positions create are spiral arms. Since the exact positions of spiral arms are highly debated, the following discussion does not use a spiral arm model. More importantly, after testing different spiral arm models, small offsets between our catalogue and the model made analysis inconclusive. Due to these observational limitations, our results and the following discussions are tentative but if present, they reveal that the organisation of gas on large scales impacts not only how clouds evolve, but where they evolve.

We weakly find on Fig.~\ref{fig:IR-MW-large} that denser environments where we detect more clouds in a smaller area, such as large scale spiral-arm-like structures, are a prerequisite for creating the large cloud population. 
Physically, this could mean that spiral arm-like locations allow some clouds (i.e., the large clouds) to grow more massive due to deeper potential wells or access to more gas. Or, due to the source blending at spiral arm locations, we are detecting more column along the line of sight over a small spatial area, resulting in higher masses and higher peak column densities. While we cannot eliminate source blending, we do minimise it by excluding regions that contained too many different distance and velocity values (See Sect \ref{sec:dist}). Furthermore, the background filtering performed on the column density maps removes material that would contribute to the source blending. Finally, since we also detect a lot the small clouds in spiral arms locations, we expect both cloud types should be impacted by source blending, meaning that differences between these two populations of clouds should be independent of the source blending. Irregardless of the dominant reason, the larger scale spiral arm structure is likely bringing gas together in such a way that we detect a significant amount of star formation from a small area that will eventually result in larger stellar associations in that small area.

Changes in clouds properties between interarm and spiral arm molecular clouds have been shown before in the Milky Way (using CO) at similar spatial resolutions \citep{sawada_structural_2012,moore_effect_2012,rigby_chimps_2019} and in nearby galaxies with well defined spiral structure at larger spatial scales \citep{colombo_pdbi_2014,faustino_vieira_molecular_2024}. In particular, \cite{sawada_structural_2012} and \cite{colombo_pdbi_2014} found two cloud populations that correlate with environment in a similar way to our work. They found that clouds in the interarm tend to be lower mass and have lower contrasts between strong and weak emission, but also indicate these clouds can exist within the spiral arms, which matches our results. Furthermore, they also found that the second population of clouds with high contrasts and higher masses primarily located in spiral arms.

To truly determine the mechanism that dominates the cloud growth of large clouds, tracing the surrounding H{\sc i} in relation to the molecular gas, surveying out to the outer galaxy to compare cloud growth in a lower density environment \citep{beuther_hiohrecombination_2016,soler_history_2020}, and using Milky Way simulations resolving down to 100~\msun\ to estimate cloud collision timescales and H{\sc i} environments (e.g., \citealt{dobbs_frequency_2015,smith_distribution_2023}) in the spiral arms are needed.

\section{Conclusions} \label{sec:conclude}

In this paper, we traced how evolved the star formation is within $\sim$12,000 molecular clouds to test how cloud properties change as star formation progresses. We identified these clouds by applying a column density cut of 1\e$^{22}$~cm$^{-2}$ \textit{Herschel}--Hi-GAL column density maps between $|\ell|<\ang{70;;}$ and $|b|<\ang{1;;}$. To measure the evolutionary state of clouds, we calculated their infrared bright fraction, \fIRB\ using \textit{Spitzer} GLIMPSE 8~\micron\ observations. This fraction corresponds to the fraction of a cloud that is 8~\micron\ against the background emission and traces the relative evolution of clouds due to the fact that 8~\micron\ emission can represent both starless, and star-forming gas when compared against a bright background 8~\micron\ source. 

\fIRB\ revealed that molecular clouds form stars simultaneously with their formation from the galactic environment. They gain mass and become denser despite their ongoing star formation. As a result, the final amount of stars that a region can form is not set purely by the initial amount of gas present in starless molecular clouds. To better view subtle secondary trends observed in cloud properties, we performed PCA. PCA revealed molecular clouds properties bifurcated down two evolutionary paths that reflected how much mass they gain as star formation progresses. Most molecular clouds (80\%) gain on average four times their original mass (termed small clouds) but the remaining 20\% percent instead gain 150 times more mass (termed large clouds). As a result, both cloud types account for $\sim$50\% of the star formation occurring within the survey area.

Both cloud types eventually stop growing and begin losing mass, but only after half their area becomes star forming. We see the mass of small clouds decrease gradually, while their peak column density remains high. The stellar feedback from the recently formed stars likely removes their more diffuse gas when their dense centre remains star forming. Large clouds rarely surpass an \fIRB\ of 0.5. We predict large clouds break up into smaller clouds due to feedback (potentially in combination with additional disruptive forces such as shear), resulting in massive evolved fragments being identified in the small cloud category.

The rate of mass gain seen in small clouds can be explained via cloud-cloud collisions, or simply accreting the surrounding lower density gas. We expect the small mass gain is caused by feedback limiting diffuse gas accretion quickly at early stages. Large clouds can achieve high mass gain due to continuous accretion from flows external to the cloud that provides a fresh supply of H$_2$ that outpaces its removal from feedback processes. Clouds with access to higher-density gas-rich environments, or higher gravitational potentials, such as spiral arms, likely provide the conditions needed for this rapid growth. We see tentative evidence for such environmental differences as we find \fIRB\ is higher and there are more large clouds in arm-like environments. However, until galactic structure and cloud distances are better constrained, we cannot yet confirm this.

Altogether, our results imply star formation is a quick but inefficient process at cloud scales due to stellar feedback limiting cloud growth. Their ability to grow in mass whilst forming stars also implies that we will under-predict the star forming ability of molecular clouds at early stages when treating them as static, isolated objects. This has far reaching implications on using a metric such as the SFE per free-fall time for cloud studies.

\section*{Acknowledgements}

We would like to thank the referee for their constructive feedback that helped improve the quality of this work. EJW acknowledges funding from the Deutsche Forschungsgemeinschaft (DFG, German Research Foundation)  -- Project-ID 138713538 -- SFB 881 (``The Milky Way System'', subproject P1). EJW and KK gratefully acknowledges funding from the DFG in the form of an Emmy Noether Research Group (grant number KR4598/2-1, PI Kreckel) and the European Research Council’s starting grant ERC StG-101077573 (“ISM-METALS"). EJW also acknowledges the support of STFC consolidated grant number ST/X001229/1. AJR acknowledges postdoctoral support from the University of Leeds. NP and AJR acknowledges the support of STFC consolidated grant number ST/N000706/1 and ST/S00033X/1. GAF gratefully acknowledges the Collaborative Research Center 1601 (SFB 1601 sub-project A2) funded by the Deutsche Forschungsgemeinschaft (DFG, German Research Foundation) – 500700252. GAF also acknowledges support from the University of Cologne and its Global Faculty programme.

\section*{Data Availability}

The data underlying this article will be shared on reasonable request to the corresponding author. 

\bibliographystyle{mnras}
\bibliography{used_ir_paper_refs5}

\appendix 

\section{Additional figures illustrating merged catalogue} \label{sec:merge_checks}

\begin{figure}  
     \centering  
     \includegraphics[width = \columnwidth]{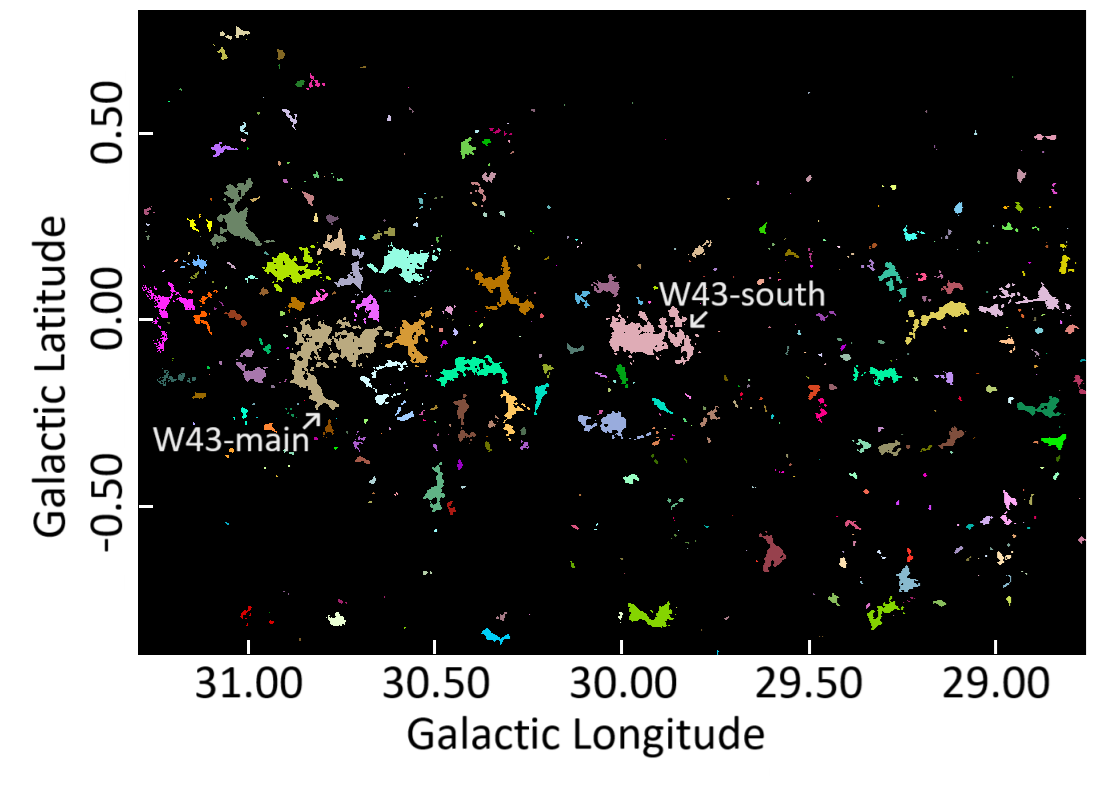}  
     \caption{Merged catalogue at bar end at $\ell=\ang{30;;}$ labelled with the location of W43.} 
    \label{fig:l30}  
\end{figure}

\begin{figure}  
\centering  
\includegraphics[width = \columnwidth]{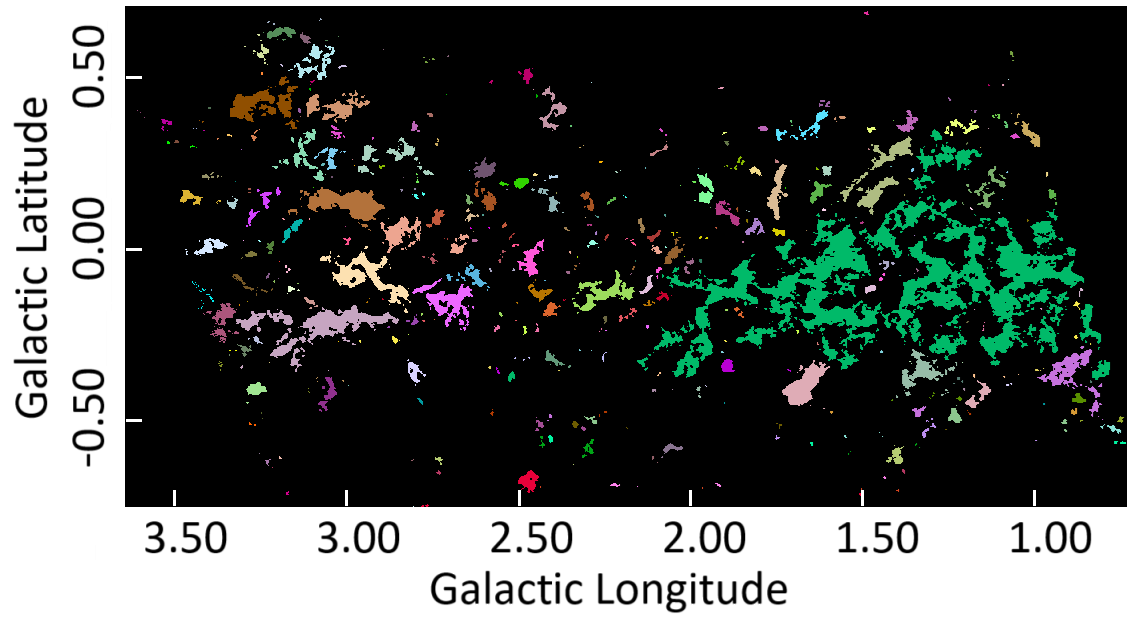}  
\caption{Merged catalogue towards the galactic centre $\ell=\ang{30;;}$.}  
\label{fig:l2}  
\end{figure}

In this appendix we show two further examples of the cloud catalogue after applying the merging algorithm at two crowded locations in addition to the example shown on Fig.~\ref{fig:24_lower}. The first example (See Fig.~\ref{fig:l30}) shows the catalogue at the bar end at $\ell=\ang{30;;}$ and the second shows the catalogue towards the galactic centre $\ell=\ang{2;;}$ (See Fig.~\ref{fig:l2}). At the bar end, objects have been successfully merged together without regions being over-merged. We note that W43-main is missing a small section to the right, though W43-south has been successfully combined. Towards the galactic centre however, we illustrate how the high column densities present and the crowding cause clouds to over-merge below $\ell=\ang{2.5;;}$. As a result, we exclude $|\ell|=\ang{2.5;;}$ from analysis.

\begin{figure*}
	\centering
		\includegraphics[width = \textwidth]{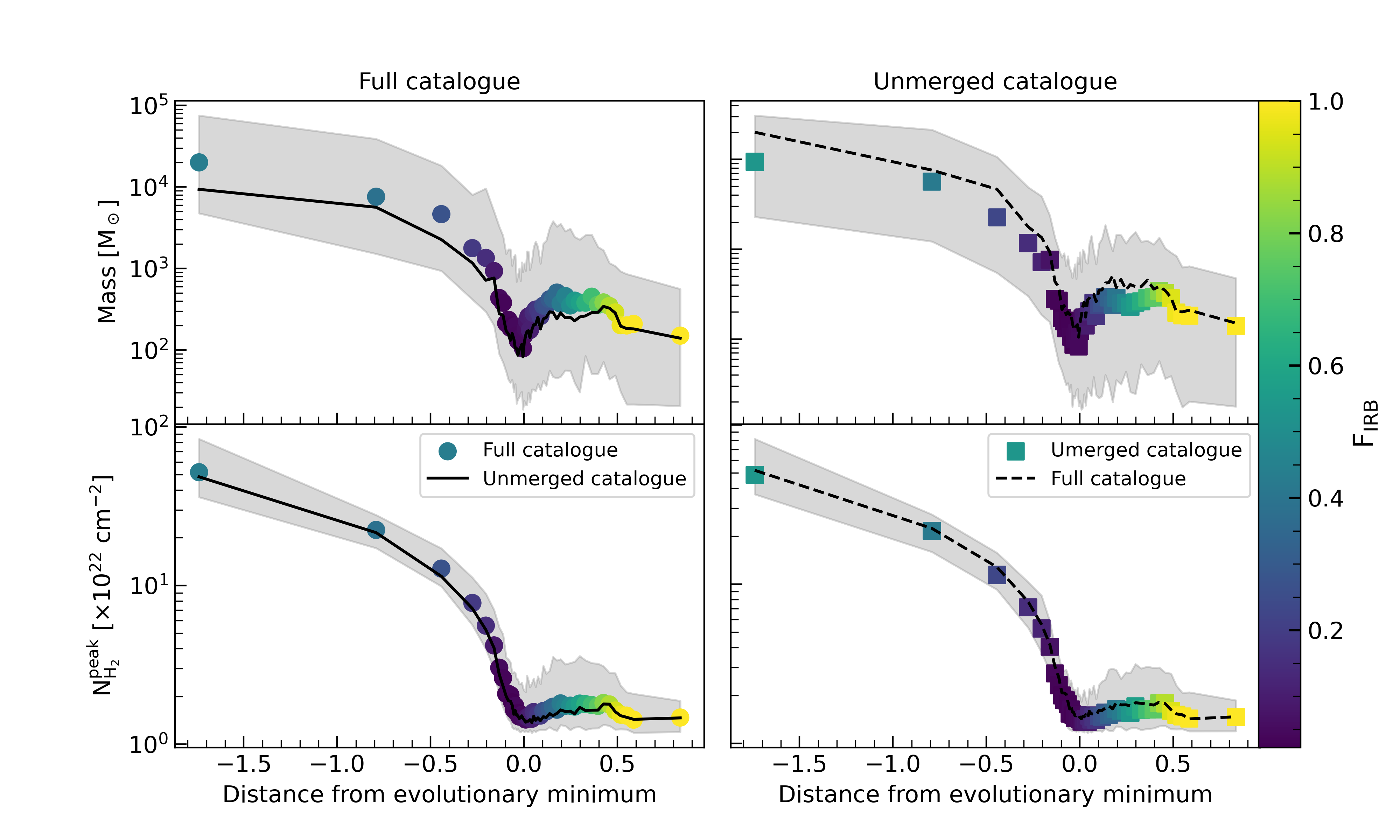}
		\caption{Median mass (top panels) and peak column density (bottom panels) as a function of their evolution. Left side (circle markers) shows the entire catalogue while the right side (square markers) shows a sub-sample of clouds that did not merge with another structure. The marker colours show the median \fIRB\ and the grey-shaded region show the 16-84\% sigma spread in the mass and peak column density. To aid comparison, solid black line of the left panel shows the values of the non-merged sample and the dash black line on the right panel shows the values of the entire catalogue.}
	\label{fig:yinc-apend}
\end{figure*} 

\section{Investigating sources of bias that impact the PCA} \label{sec:checks}

We check if the bifurcation of cloud properties as a function of evolution could be caused by the cloud merging algorithm we use in Sec.~\ref{sec:cloud_id}. Potentially, the large clouds might preferentially contain clouds that merge, causing the mass increase that we see (a form of source crowding). To test this, we remade Fig.~\ref{fig:yinc} but only include clouds that did not merge with another cloud. To help with the comparison, we plot the full data set (left panel) and the non-merged data set (right panel) in Fig.~\ref{fig:yinc-apend}. The black line on each panel show the values of the other panel to aid the comparison.  We find that the median mass and the peak column density values are nearly identical for the merged and non-merged samples and we still see two evolutionary paths. Therefore, the cloud merging process does not cause there to be two cloud populations.

\begin{figure}
	\centering
		\includegraphics[width = \columnwidth]{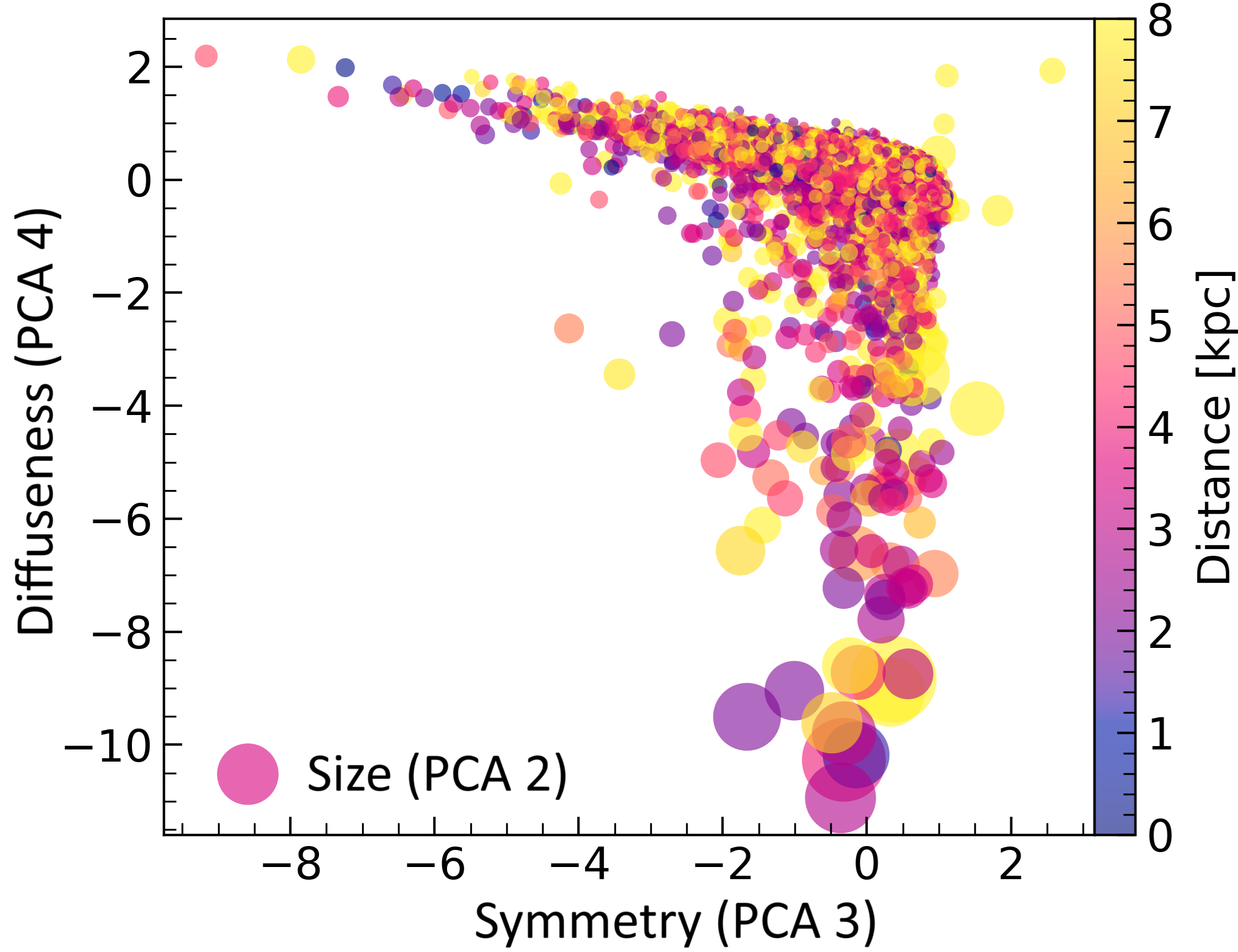}
		\caption{Score plot of the cloud catalogue using their principal components compared to their assigned distance. The diffuseness axis (PC4) is represented using the y axis, the symmetry axis (PC3) is represented using the x axis, the size of the markers shows the size of the clouds and the marker colours show the heliocentric distance assigned to each cloud.
		}
	\label{fig:pca4-dist}
\end{figure}

\begin{figure}
	\centering
		\includegraphics[width = \columnwidth]{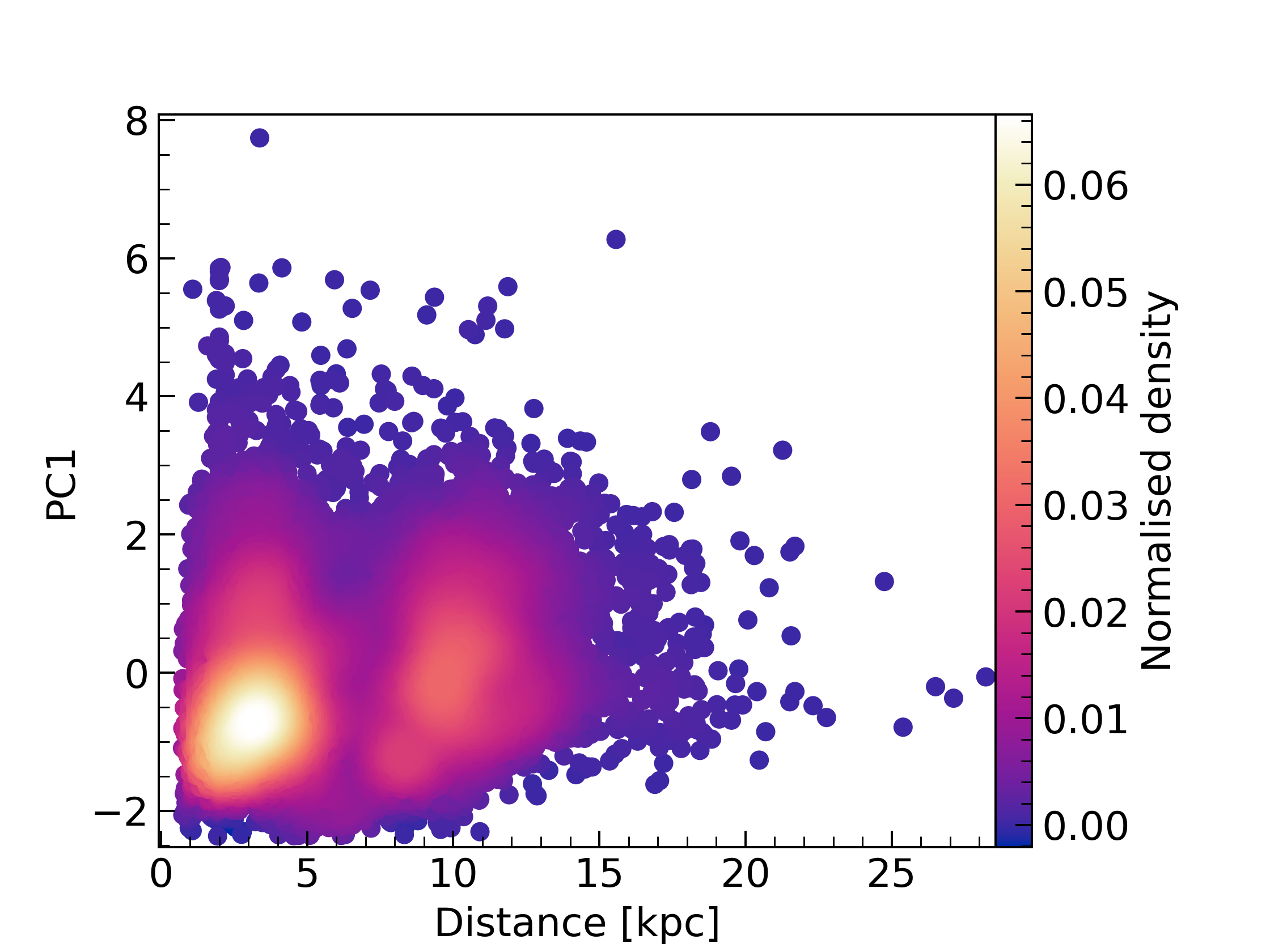}
		\caption{PC1 against assigned heliocentric distance. Colour of markers indicate the 2D histogram density of points.}
	\label{fig:pca1-dist}
\end{figure}

We next check if the potential distance bias (discussed in Sect. \ref{sec:contrast}) is present in the PCA. Due to distance, we are more likely to find both higher mass clouds, and higher \fIRB\ fractions clouds as a function of distance (see Sect \ref{sec:contrast} for explanation of \fIRB\ bias). Mass bias is a result of resolution. Clouds which are further away will have less detectable substructure and are more likely to blend neighbouring clouds into one object. Distance limiting the sample minimises both \fIRB\ bias and mass bias, but removes most of the clouds on arm environments and limits our discussion. The distance bias can therefore affect the PCA, and therefore the azmuthial trends we find in the two cloud types. Using Kendall's $\tau$ correlation coefficient, we check for distance bias. 
\begin{table}
    \centering
    \begin{tabular}{l|cc}
                      & $\tau$ & p-value \\
    \hline
     PC1 (Evolution)  & 0.13   & 7.5\e$^{-107}$ \\
     PC2 (Size)       & -0.10  & 5.7\e$^{-59}$  \\
     PC3 (Symmetry)   & 0.01   & 0.31           \\
     PC4 (Diffuseness) & 0.11   & 1.2\e$^{-70}$  \\
    \end{tabular}
    \caption{Table showing how each principle component correlates with distance. We find weak or negligible correlation for each component}
    \label{tab:kendall}
\end{table}
We find each component has a correlation coefficient of $\lesssim|0.1|$ with p-values of 0, except for PC3, which has a coefficient of 0.01 and p-value of 0.31 indicating the correlation is indistinguishable from random chance. We show their exact statistics in Table \ref{tab:kendall}. For visual confirmation of these correlations we replot Fig.~\ref{fig:pca4} on Fig.~\ref{fig:pca4-dist} and colour the markers using their distance to check if and how distance correlates with the components. We see there is no obvious correlation with distance. We also plot PC1 against distance on Fig.~\ref{fig:pca1-dist} and can see there a only a weak positive correlation. We are thus confident that distance does not play a dominant role in determining the PCA components, and since the PCA forms the majority of our analysis and discussion, we are also confident that distance bias is not responsible for the conclusions we draw.

\begin{table*}

    \centering
    \begin{tabular}{c|ccccc}

                   & Peak column & Mass  & Temperature & \fIRB & Aspect ratio \\
                   & density     &       &             &       &              \\
\hline
Original value     & 0.301       & 0.092 & 0.679       & 0.661 & -0.060       \\
Median             & 0.303       & 0.093 & 0.678       & 0.660 & -0.059       \\
Mean               & 0.307       & 0.106 & 0.676       & 0.657 & -0.058       \\
Standard deviation & 0.023       & 0.042 & 0.012       & 0.014 & 0.012        \\
16th percentile	   & 0.294       & 0.089 & 0.675       & 0.656 & -0.070       \\
84th percentile    & 0.313       & 0.123 & 0.680       & 0.663 & -0.047       \\
    \end{tabular}
    \caption{Statistics of how the normalised eigenvectors change after randomly swapping 20\% of the distances to their other distance solution 10,000 times and re-performing PCA.}
    \label{tab:avg_eigvec}
\end{table*}

\begin{table*}
    \centering
    \begin{tabular}{c|ccccc}
                   & PC1   & PC2   & PC3      & PC4   & PC5      \\
                   & (Evolution) & (Size) & (Symmetry) & (Diffuseness) &  \\
\hline
Original value     & 0.302 & 0.217 & 0.199    & 0.174 & 0.107    \\
Median             & 0.302 & 0.218 & 0.200    & 0.173 & 0.107    \\
Mean               & 0.303 & 0.219 & 0.199    & 0.172 & 0.107    \\
Standard deviation & 0.001 & 0.006 & 0.000(6) & 0.007 & 0.000(7) \\
16th percentile    & 0.301 & 0.216 & 0.199    & 0.171 & 0.107    \\
84th percentile	   & 0.304 & 0.220 & 0.200    & 0.173 & 0.108    \\
    \end{tabular}
    \caption{Statistics of how the normalised eigenvalues change after randomly swapping 20\% of the distances to their other distance solution 10,000 times and re-performing PCA.}
    \label{tab:avg_eigval}
\end{table*}

\begin{figure*}
	\centering
		\includegraphics[width = \textwidth]{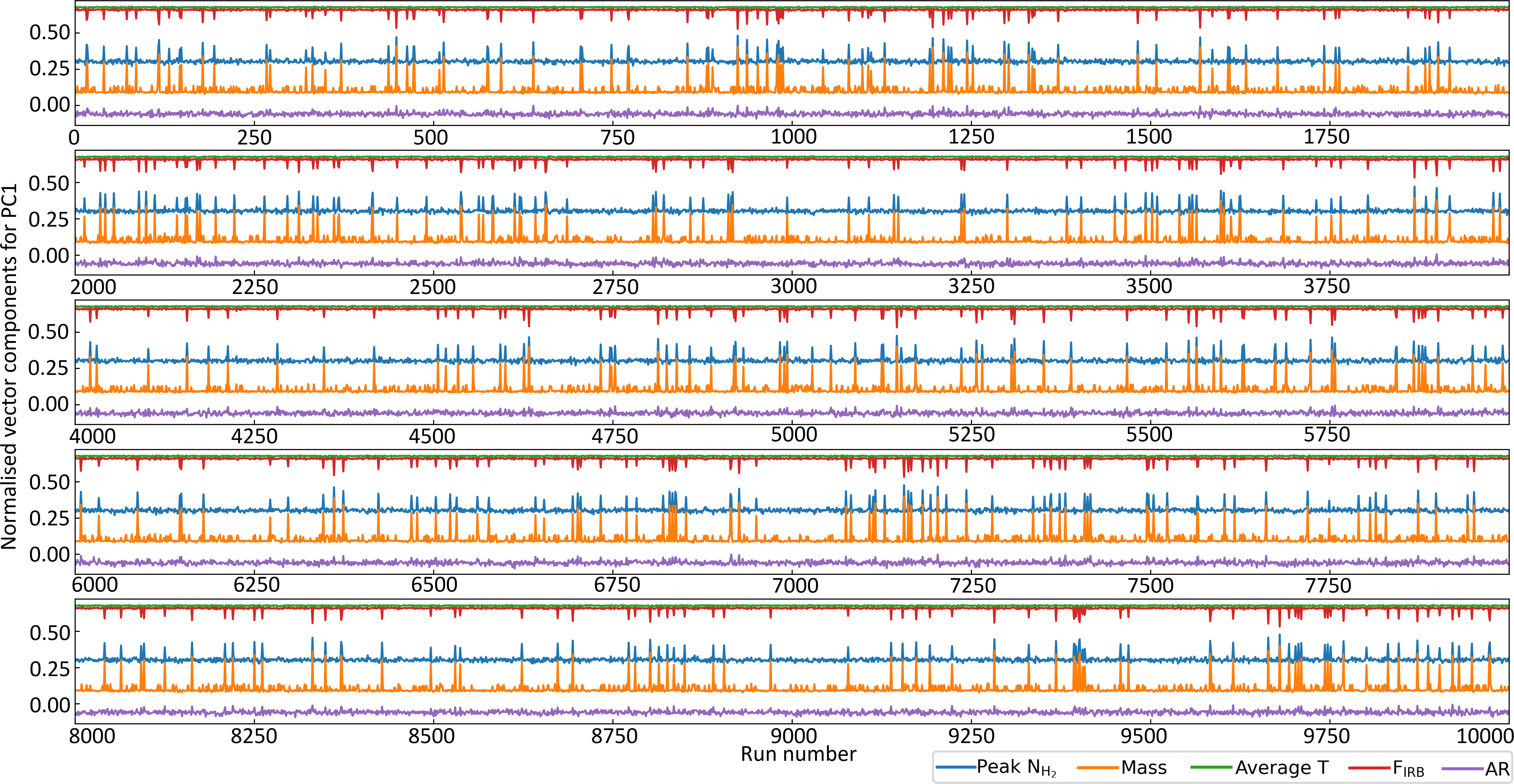}
		\caption{How the normalised eigenvectors change after randomly swapping 20\% of the distances to their other distance solution 10,000 times and re-performing PCA. Blue line shows the peak column density, orange shows the mass, green shows the average temperature, red shows \fIRB, and purple shows the aspect ratio.
		}
	\label{fig:eigvec}
\end{figure*} 

\begin{figure*}
	\centering
		\includegraphics[width = \textwidth]{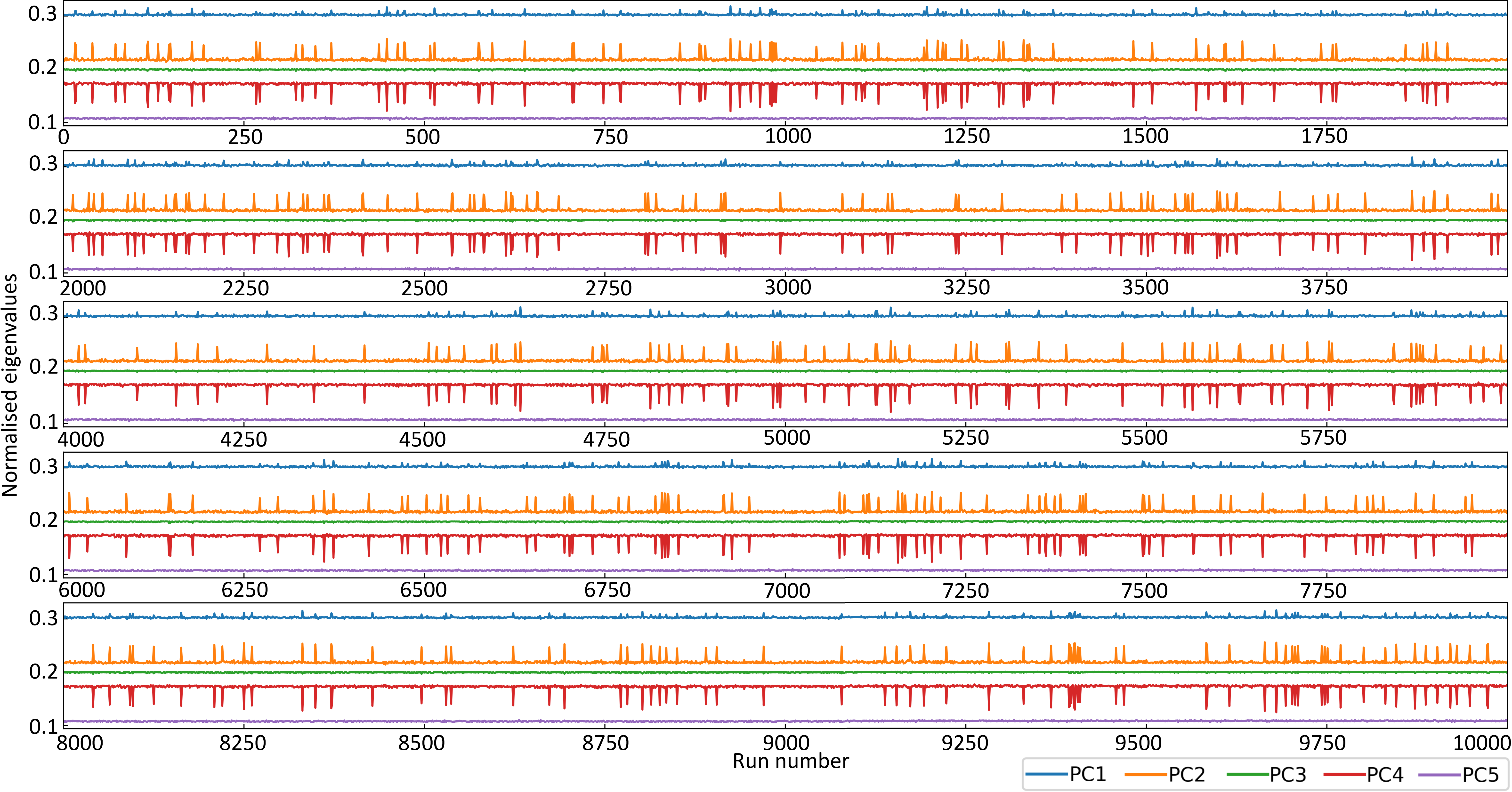}
		\caption{How the normalised eigenvalues change after randomly swapping 20\% of the distances to their other distance solution 10,000 times and re-performing PCA. Blue line shows PC1, orange shows PC2, green shows PC3, red shows PC4, and purple shows PC5.}
	\label{fig:eigval}
\end{figure*} 

We finally check how robust the PCA components are to incorrect distance assignments caused by the near-far ambiguity. In Sect \ref{sec:dist}. we preferentially select near distances when multiple combined structures had matching velocities with both near and far assignments. Understandably, there will be incorrect assignments in our catalogue. Additionally, we find that the number of near-far assignments in the original Hi-GAL clump catalogue we use deviates from other clump based studies (such as \citealt{urquhart_atlasgal_2018}) by 20\% on the near-far assignment. To understand how sensitive our results are to incorrect assignments, we randomly swap 20\% of the distances to their near or far counterpart then recalculate the PCA 10,000 times. We then quantify the changes in the PCA by looking at statistics (i.e., mean) of how the eigenvectors and values change in Table \ref{tab:avg_eigvec} and Table \ref{tab:avg_eigval} respectively. We find there is very little spread and the average values are nearly identical to the original PCA vectors we find, providing us confidence that our results and discussions are not impacted by incorrect near-far distance assignments. Furthermore we find that the PCA solution changes slightly only 4\% of the time based of the variance between the data with the average values. To illustrate these locations, we plot the values of the normalised eigenvectors and values for all 10,000 tests on Fig.~\ref{fig:eigvec} and Fig.~\ref{fig:eigval} respectively. The `spikes' on the figures show where the changed distance assignment impacted the PCA solution.

\section{Face-down Milky Way plots} \label{sec:mw-birdseye}

\begin{figure*}
	\centering
		\includegraphics[width = 0.95\textwidth]{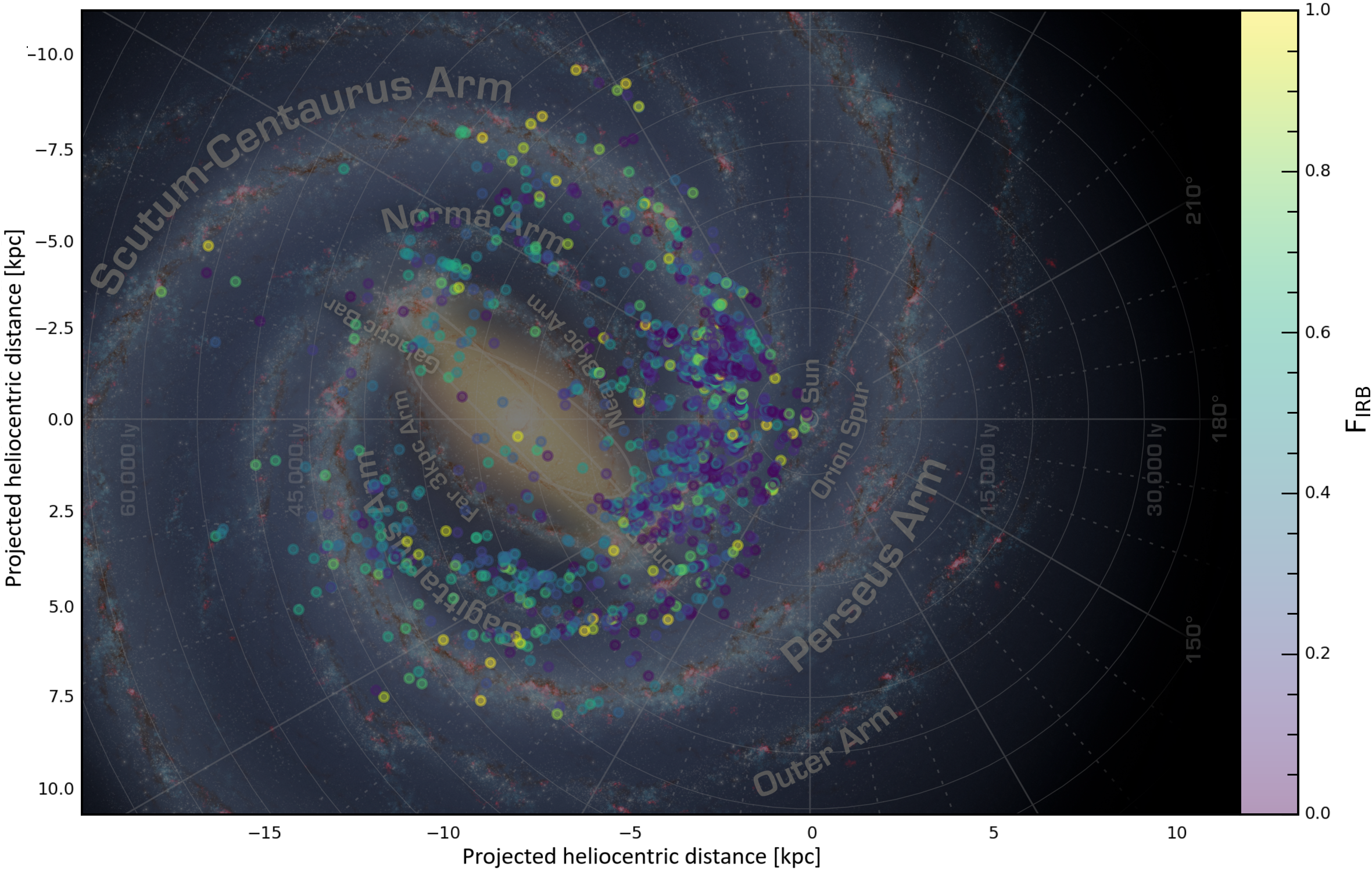}
		\caption{\fIRB\ of \textit{large clouds} only over-plotted over the Milky Way. Marker colours represent the \fIRB\ of each cloud shown.}
	\label{fig:IR-MW-large}
\end{figure*} 

\begin{figure*}
	\centering
		\includegraphics[width = 0.95\textwidth]{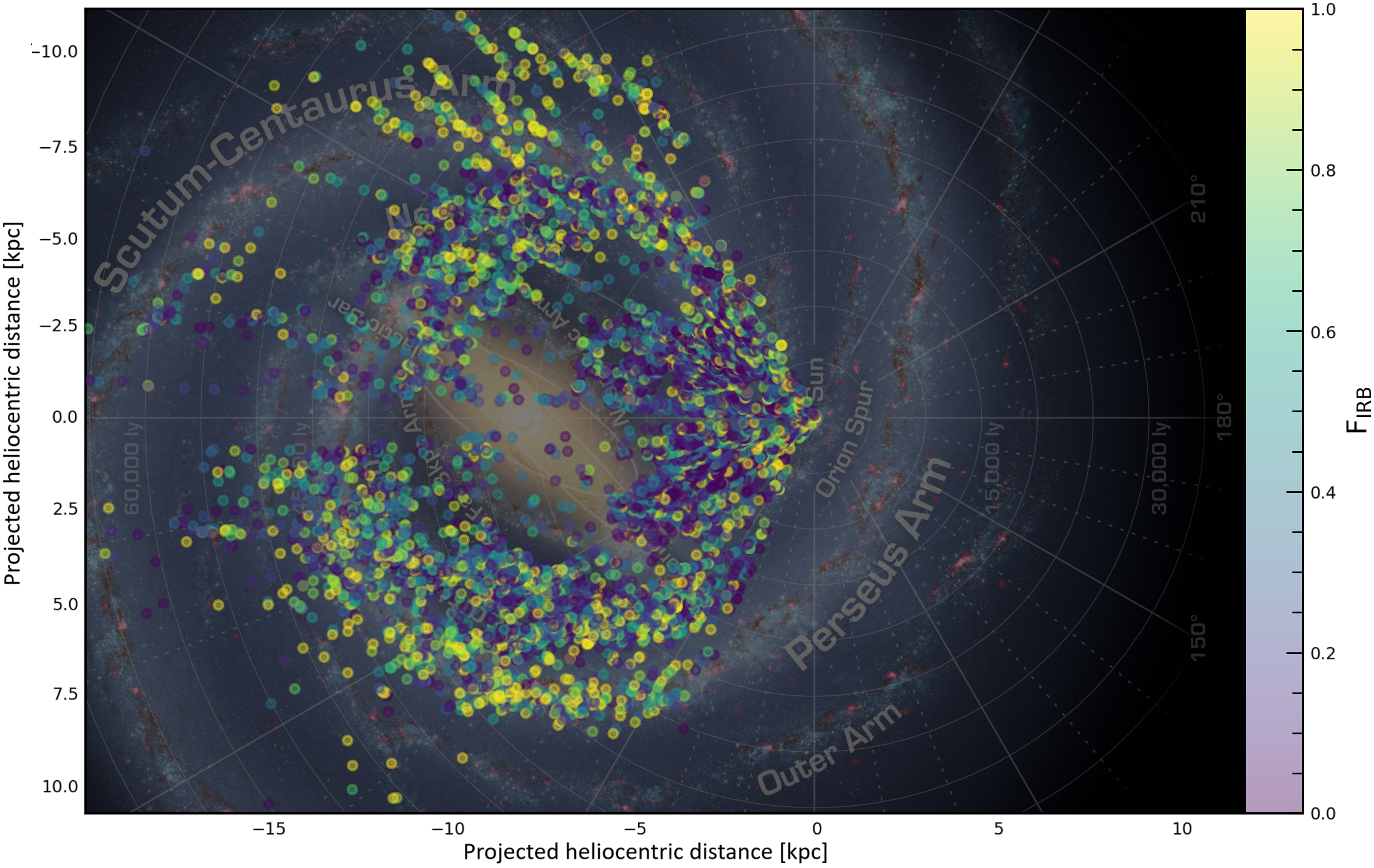}
		\caption{Same as Fig.~\ref{fig:IR-MW-large} but for \textit{small clouds} only.}
	\label{fig:IR-MW-small}
\end{figure*}

This section presents the top-down view of difference sub samples of the cloud catalogue over an illustration of the Milky Way structure. We show the top down view of the large and small cloud sample in Fig.~\ref{fig:IR-MW-large} and Fig.~\ref{fig:IR-MW-small} respectively.

\bsp	
\label{lastpage}
\end{document}